\documentclass{aa}

\usepackage{epsfig,graphicx,natbib}
\usepackage{longtable}
\usepackage{amssymb}
\usepackage{amsmath}
\usepackage{mathrsfs} 
\usepackage{color}
\usepackage{subcaption}
\usepackage{booktabs}
\usepackage{xspace}
\usepackage{float}
\usepackage{longtable}
\usepackage{breqn}
\usepackage{algpseudocode}
\usepackage{algorithm}
\usepackage{hyperref}

\usepackage{todonotes}
\usepackage{adjustbox}

\newcommand{\norm}[1]{\lVert #1 \rVert}

\newcommand{\argmin}{\mathop{\mathrm{arg\:min}}}

\newcommand{\gapprox} {\, \lower3pt\hbox{$\sim$}\llap{\raise2pt\hbox{$>$}}\,}

\begin{document}

\title{Identifying synergies between VLBI and STIX imaging}

\author{ Hendrik Müller \inst{1,2}, Paolo Massa \inst{3}, Alejandro Mus \inst{4,5}, Jong-Seo Kim \inst{1}, Emma Perracchione \inst{6}
}

\institute{
   Max-Planck-Institut für Radioastronomie, Auf dem Hügel 69, D-53121 Bonn (Endenich), Germany \\ \email{hmueller@mpifr-bonn.mpg.de} 
   \and 
   National Radio Astronomy Observatory, 1003 Lopezville Rd, Socorro, NM 87801, USA
   \and
   Department of Physics \& Astronomy, Western Kentucky University, 1906 College Heights Blvd., Bowling Green, KY, 42101, USA
   \and
    Departament d’Astronomia i Astrof\'isica, Universitat de Val\`encia, C. Dr. Moliner 50, 46100 Burjassot ,Val\`encia, Spain
  \and
  Observatori Astron\`omic, Universitat de Val\`encia, Parc Cient\'ific, C. Catedr\`atico Jos\'e Beltr\'an 2, 46980 Paterna, Val\`encia, Spain 
  \and
  Dipartimento di Scienze Matematiche “Giuseppe Luigi Lagrange”, Politecnico di Torino, Corso Duca degli Abruzzi, 24, 10129, Torino, Italy
}

\date {Received  / Accepted}

\authorrunning{Müller, Massa, Mus, Kim, Perracchione}

\abstract
%context heading (optional)
{Reconstructing an image from noisy, sparsely sampled Fourier data is an ill-posed inverse problem that occurs in a variety of subjects within science, including the data analysis for Very Long Baseline Interferometry (VLBI) and the Spectrometer/Telescope for Imaging X-rays (STIX) for solar observations. The need for high-resolution, high-fidelity imaging fosters the active development of a range of novel imaging algorithms in a variety of different algorithmic settings. However, despite this ongoing parallel developments, synergies remain unexplored.}
% aims heading (mandatory)
{We study for the first time the synergies between the data analysis for the STIX instrument and VLBI. Particularly, we compare the methodologies that have been developed in both fields and evaluate their potential. In this way, we identify key trends in the performance of several algorithmic ideas and draw recommendations for the future spending of resources in the study and implementation of novel imaging algorithms.}
% methods heading (mandatory)
{To this end, we organized a semi-blind imaging challenge with data sets and source structures that are typical for sparse VLBI, specifically in the context of the Event Horizon Telescope (EHT), and for STIX observations. 17 different algorithms from both communities, from 6 different imaging frameworks, participated in the challenge, marking this work the largest scale code comparisons for STIX and VLBI to date.}
% results heading (mandatory)
{Strong synergies between the two communities have been identified, as can be proven by the success of the imaging methods proposed for STIX in imaging VLBI data sets and vice versa. Novel imaging methods outperform the standard CLEAN algorithm significantly in every test-case. Improvements over the performance of CLEAN make deeper updates to the inverse modeling pipeline necessary, or consequently replacing inverse modeling with forward modeling. 
Entropy-based methods and Bayesian methods perform best on STIX data. The more complex imaging algorithms utilizing multiple regularization terms (recently proposed for VLBI) add little to no additional improvements for STIX, but outperform the other methods on EHT data, which correspond to a larger number of angular scales.}
% conclusions heading (optional)
{This work demonstrates the great synergy between the STIX and VLBI imaging efforts and the great potential for common developments. The comparison identifies key trends on the effectivity of specific algorithmic ideas for the VLBI and the STIX setting that may evolve into a roadmap for future developments.}


\maketitle

\section{Introduction}
Inverse problems are a class of problems for which the causal factors that lead to certain observables are recovered, opposed to a forward problem in which the observables are predicted based on some initial causes. A common difficulty when solving inverse problems is \textit{ill-posedness}, i.e. the direct (pseudo-)inverse (if it exists and is single-valued) is unstable against observational noise \citep{Hadamard1953}. Therefore, in order construct reliable approximations of the (unknown) solution, additional prior information has to be encoded in the reconstruction process, and this procedure is called \textit{regularization} \citep{Morozov1967}. Ill-posed inverse problems arise in a variety of settings, for example for medium scattering experiments \citep[e.g.][]{Colton2013}, medical imaging \citep[e.g. see the review][]{Spencer2020}, or microscopy. In an astrophysical context, ill-posed inverse problems include among others for Ly$\alpha$ forest tomography \citep{Lee2018, Mueller2020, Mueller2021}, lensing \citep[e.g. see the review][]{Mandelbaum2018}, or (helio-)seismology \citep{Gizon2010}.

A specific class of inverse problems is the reconstruction of a signal from a sparsely-undersampled Fourier domain. This problem formulation applies interdisciplinary to radio interferometry \citep{Thompson2017}, magnetic resonance tomography \citep{Spencer2020}, as well as to solar hard X-ray imaging \citep{pianabook}. Although, the fundamental problem formulation is similar, the degree of undersampling, the spatial scales and features of the scientific targets, the calibration effects, and noise corruptions differ. Therefore, the scientific communities developed independently from each other multiple regularization methods specifically tailored to the needs of the respective instruments. 
That resulted at some occasions in duplicated, parallel developments \citep[e.g. of modern maximum-entropy methods: ][]{Chael2016, Massa2020, Mus2023}, while, at other instances, to complementary, mutually inspiring, and interdisciplinary efforts \citep[e.g. multiobjective optimization: ][]{Mueller2023}. 

Given that image reconstruction from solar X-ray imaging and radio interferometry data is a rather challenging problem without unique solutions, there is no algorithm which is consistently optimal under every circumstance. Therefore, research on imaging algorithms for solar X-ray imaging and for radio interforemtry, and particularly Very Long Baseline Interferometry (VLBI), are currently active fields of science: see e.g. recent works for VLBI by \citet{Issaoun2019, Broderick2020, Broderick2020b, Arras2022, Sun2022, Tiede2022, Mueller2022, Mueller2022b, Mueller2022c, Mueller2023, ngehtchallenge, Mus2023, Chael2023, Kim2024} and for solar X-ray imaging by \citet{Massa19, Massa2020, Siarkowski2020, Perracchione2023}. Here we compare the data reduction methods that were recently proposed for solar X-ray imaging with the recent developments from the field of VLBI and identify their mutual potential to be applied in the respective opposite field. This work constitutes and contributes to a roadmap for future developments in imaging and spending of scientific resources.

In VLBI (and radio-interferometry in general), multiple different-placed antennas observe the same source at the same time. The correlation product of the signals recorded by an antenna pair in the array during an integration time approximately determines a Fourier component (i.e. a \textit{visibility}) of the true sky brightness distribution \citep{Thompson2017}. The Fourier frequency is determined by the baseline separating the two antennas of one antenna pair projected onto the sky plane. This is described by the van Cittert-Zernike theorem \citep{vanCittert1934, Zernike1938}. The angular frequency domain (often referred to as the $(u,v)$-domain) gets ``filled'' due to the rotation of the Earth with respect to the source on the sky. However, due to limited numbers of antennas in the array, the $(u,v)$-domain is only sparsely covered by measurements. The set of frequencies measured by the antenna array is called $(u,v)$-coverage. The image of the radio source is then derived from the visibility set through a Fourier inversion process \citep{Thompson2017}. Moreover, VLBI often deals with challenging calibration issues, among with scale-dependent thermal noise, particularly at mm-wavelengths \citep{Janssen2022}. At imaging stage, these are typically factored out in station-based gains.  

\begin{figure*}
    \centering
    \includegraphics[width=0.7\textwidth]{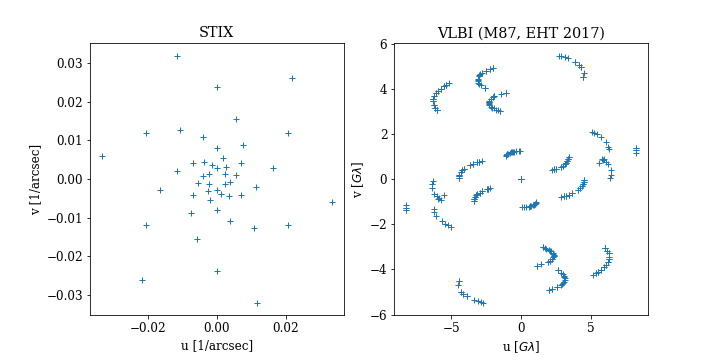}
    \caption{Exemplary comparison of the STIX $(u,v)$-coverage (\textit{left panel}) and of the EHT one \citep[\textit{right panel};][]{eht2019a, eht2019d}. In both cases the Fourier domain is undersampled, although the degree of undersampling is more enhanced in the STIX case.}
    \label{fig: comparison}
\end{figure*}

The Spectromenter/Telescope for Imaging X-rays \citep[STIX:][]{Krucker2020} is the instrument of the ESA Solar Orbiter mission \citep{2020A&A...642A...1M} dedicated to the observation of the X-ray radiation emitted by solar flares. The telescope provides diagnostics on the temperature of the plasma and on the flare-accelerated electrons by observing the corresponding X-ray radiation emitted by thermal and non-thermal bremsstrahlung. STIX modulates the flaring X-ray radiation by means of pairs of grids. The X-ray flux transmitted by each grid pair creates a Moir\'e pattern, whose intensity is measured by a coarsely pixelated detector. The measurements provided by the detector allow for the determination of amplitude an phase for a single visibility of the flaring X-ray source. Thus, similarly to VLBI, the data provided by STIX is a set of visibilities that can be used for image reconstruction of the flare X-ray emission. However, there are some differences between STIX and VLBI. Due to the smaller number of visibilities (30 for STIX vs. $\gapprox$150 for VLBI; there is a comparison of the $(u,v)$-coverages in Fig. \ref{fig: comparison}) the achievable dynamic range for STIX is \(\sim\)10. The synergy is therefore greatest to VLBI snapshot imaging. Moreover, while it has become more common in mm-VLBI to do the imaging without phase-information \citep{Chael2018, Mueller2022}, for STIX well-calibrated visibility-phases are available. Nevertheless, at the beginning of the STIX visibility phase calibration process, the problem of image reconstruction from visibility amplitudes alone has been addressed for STIX \citep{Massa2021}.

Imaging routines that were proposed for STIX include Back-Projection \citep{Hurford2002}, CLEAN \citep{Hogbom1974} and its recent unbiased version U-CLEAN \citep{Perracchione2023}, Expectation Maximization \citep{Massa19,Siarkowski2020}, the Maximum Entropy Method MEM\_GE \citep{Massa2020}, and the parametric forward-fitting method VIS\_FWDFIT \citep{Volpara2022}. In radio astronomy the classical de-facto standard is CLEAN and its many variants \citep{Hogbom1974, Schwarz1978, Schwab1984}. CLEAN has been extended to the multiscalar domain, primarily to adapt to extended emission \citep{Wakker1988, Starck1994, Bhatnagar2004, Cornwell2008, Rau2011, Offringa2017} or to allow super-resolution within CLEAN \citep{Mueller2022b}. Maximum Entropy methods have been historically suggested as an alternative \citep[e.g.][]{Cornwell1985}, and have been studied in the context of compressive sensing since then \citep[among others][]{Pantin1996, Wiaux2009, Garsden2015}. A wide variety of additional algorithms have been recently proposed for radio interferometry including Bayesian algorithms \citep[e.g.][]{Arras2019, Arras2021, Kim2024}, deep learning \citep{Aghabiglou2022, Dabbech2022, Terris2023, Wilber2023}, and compressive sensing \citep{Mueller2022, Mueller2022b, Wilber2023b}. For this work, we focus on sparse VLBI arrays such as the Event Horizon Telescope (EHT) due to the similarity of the undersampling and dynamic range to the STIX instrument. Specifically for the EHT, novel, super-resolving imaging algorithms have been developed recently in the context of Regularized Maximum Likelihood (RML) methods \citep{Honma2014, Akiyama2017, Akiyama2017b, Chael2016, Chael2018, Mueller2022, Mueller2022c}, Bayesian algorithms \citep{Broderick2020, Broderick2020b, Tiede2022} and multiobjective optimization \citep{Mueller2023, Mus2023b}.

For this work we are specifically focusing on two research questions: First, are there strong synergies between the STIX data analysis and VLBI imaging and room for mutual algorithmic development? To this end we investigate the effectivity of VLBI data analysis methods for STIX and evaluate quantitatively whether the recently proposed VLBI imaging frameworks are interesting for STIX. Vice versa, we test STIX algorithms in frontline VLBI settings. This is realized in a semi-blind data analysis challenge. 

Second, which of the numerous regularization frameworks (e.g. inverse modeling, RML, compressive sensing, maximum entropy, Bayesian algorithms, evolutionary methods) are most promising and worthy to invest resources for further development in the future? We achieve conclusive hints in this regard by basing our data analysis challenge on a wide range of methods, consisting of submissions by 17 different methods. 

The rest of the paper is structured as follows. We recall the basic introduction to the VLBI and STIX imaging problem in Sec. \ref{sec: theory}. We present the used imaging methods in Sec. \ref{sec: imaging}. We present the outline of the data analysis challenge, evaluation metrics and finally our reconstruction results in Sec. \ref{sec: challenge}. We test selected algorithms on real STIX data in Sec. \ref{sec: real_data}. We present our conclusions in Sec. \ref{sec: discussion}.

\section{Theory} \label{sec: theory}
In this section we present the basic notions of VLBI and STIX imaging. The concepts here presented are the bases for understanding the image reconstruction problem.

\subsection{VLBI}
The correlation product of an antenna pair in a VLBI array is approximately the Fourier transform of the true sky brightness distribution. This is described by the van-Cittert-Zernike theorem:
\begin{align}
    \mathcal{V} (u, v) = \iint I(l, m) e^{-2 \pi i (l u + m v)} \,dl\, dm \,.  \label{eq: vis}
\end{align}
where $I(l,m)$ is the sky brightness distribution, and $l,m$ are direction cosines. We typically ignore $w$-projection terms due to small field of view for VLBI. The observables $\mathcal{V}(u,v)$ are called visibilities with respect to harmonics $u,v$, determined by the baselines separating the antennas projected on the sky plane. When we produce an image in VLBI we are trying to recover the image intensity $I(l,m)$ from the measured visibilities. Since, the Fourier domain (i.e. the $(u,v)$-domain) is only sparsely sampled by observations, VLBI imaging constitutes an ill-posed inverse problem. The problem is further complicated by calibration effects and thermal noise. Particularly, the observed visibilities for an antenna pair $i,j$ at a time $t$ are related to the model visibilities by the relation:
\begin{align}
    V(i,j,t) = g_i g_j^* \mathcal{V}(i,j,t) + N_{i,j}\,, 
\end{align}
where $N_{i,j}$ is Gaussian thermal noise with an unknown correlation structure, and $g_i$ is complex valued gain factors specific to antenna $i$. The gain factors typically vary over the time of an observation. While most VLBI algorithms were developed and are applied in the context of gain-corrupted data sets, e.g. by alternating imaging with self-calibration loops \cite[hybrid imaging][]{Readhead1978, mus22} or by closure only imaging \citep{Chael2018, Mueller2022, Mueller2023}, we will focus in this work on dealing with undersampling and noise corruption issues and ignore the need for self-calibration in VLBI.

\subsection{STIX}
The STIX instrument contains 30 sub-collimators, i.e., units consisting of a grid pair and a detector mounted behind it. The period and orientation of the front and of the rear grid in each pair are chosen in such a way that the transmitted X-ray flux creates a Moiré pattern on the detector surface with period equal to the detector width \citep[e.g.,][]{hurford2013x,prince1988gamma}. STIX Moiré patterns encode information on the morphology and location of the flaring X-ray source. Therefore, from measurements of the transmitted X-ray flux performed by the detector pixels, it is possible to compute the values of amplitude and phase of 30 complex visibilities \citep{Massa2023}.The imaging problem for STIX can then be described by the following equation
\begin{equation}\label{eq: STIX vis}
\mathcal{V}(u_j,v_j) = \iint I(x,y) e^{2\pi i(xu_j+yv_j)}\,dx\,dy ~,
\end{equation}
where $j=1,\dots,30$ is the sub-collimator index, $\mathcal{V}(u_j,v_j)$ is experimental visibility corresponding to the $(u_j,v_j)$ angular frequency, and $I(x,y)$ is the angular distribution of the intensity of the X-ray source. Note that, differently from the VLBI case, in the STIX case there is a plus sign inside the complex exponential that defines the Fourier transform (cf. Eqs. \eqref{eq: vis} and \eqref{eq: STIX vis}). Further, the angular frequencies sampled by STIX are only determined by geometric properties of the instrument hardware \citep{Massa2023}; therefore, they are the same for every recorded event.

% STIX measures a number of visibilities which is at least an order of magnitude lower compared to that measured by VLBI, and enhances the ill-posedness of the imaging problem for the X-ray imager compared to that of the radio domain. 
STIX measures a number of visibilities which is at least an order of magnitude lower compared to that measured by VLBI. Therefore, the ill-posedness of the imaging problem for the X-ray imager is more enhanced compared to that of the radio domain. However, the great advantage of STIX is that the data calibration (in particular, the visibility phase calibration) depends only on the geometry of the sub-collimator grids and is therefore stable in time \citep{Massa2023}.

\section{Imaging Methods} \label{sec: imaging}
In this paper we compare the reconstructions from a variety of algorithms and algorithmic frameworks. We will briefly explain each of these frameworks in the following subsections. A tabular overview of the algorithms, their optimization details, their basic ideas, regularization properties, advantages and disadvantages is presented in Appendix \ref{app: imaging}.

\subsection{CLEAN}
CLEAN and its many variants \citep{Hogbom1974, Wakker1988, Schwarz1978, Bhatnagar2004, Cornwell2008, Rau2011, Offringa2017, Mueller2022b} are the de-facto standard imaging methods for VLBI and STIX. CLEAN reformulates the imaging problem as a deconvolution problem by means of the dirty beam $B^D$ and dirty map $I^D$. The former is the instrument Point Spread Function (PSF), while the latter is the inverse Fourier transform of the all measured visibilities sampled on an equidistant grid, filled by zeros for every non-measured Fourier coefficients (i.e. for gaps in the $(u,v)$-coverage), and reweighted by the noise ratio (natural weighting) or the power within a gridding cell (uniform weighting) or a combination of both \citep{Briggs1995}. In this way, CLEAN solves the problem:
\begin{align}
I^D = B^D * I.
\end{align}
In classical CLEAN \citep{Hogbom1974}, this problem is solved iteratively in a greedy matching pursuit process. In a user-defined search window, we search for the maximal peak in the residual, store the position and strength of the peak in a list of CLEAN components, shift the dirty beam to the position of the peak, and subtract a fraction of the beam (determined by the CLEAN gain) from the residual. This procedure is repeated with the new residual until the residual is noise-like, i.e. we have computed the approximation:
\begin{align}
    I^D \approx B^D * \left( \sum_i a_i \delta(l_i, m_i) \right),
\end{align}
with $a_i$ being the intensity of the $i$-th CLEAN component, and $l_i, m_i$ the coordinates of its location. Finally, denoting with $B^C$ the CLEAN beam, i.e. a Gaussian beam that is fitted to the central lobe of the dirty beam, the final CLEAN image is computed by convolving the CLEAN components with $B^C$:
\begin{align}
I^C = B^C * \left( \sum_i a_i \delta(l_i, m_i) \right).
\end{align}
It is also standard to add the last residual to the reconstructed image to account for any non-recovered flux.

While CLEAN remains in use, mainly because it is straightforward and interactive, it remains fairly limited \citep[e.g. see the recent summaries on the limitations of CLEAN][]{Pashchenko2023, Mueller2023b}. We shall summarize some of the main limitations of CLEAN here. The representation of the image by a sample of point sources is not a reasonable description of the physical image, particularly when representing extended emission. This issue has two important consequences. First, the model that is fitted to the data (i.e. the CLEAN components) and the final image (the CLEAN components convolved with the clean beam) are not the same. Particularly, for high dynamic range imaging, one therefore self-calibrates the image to a model that was deemed physically unfeasible. Second, the representation of the image by CLEAN components makes the use of a final convolving necessary, hence limiting the effective resolution. Moreover, the value of the Full Width at Half Maximum (FWHM) of the Gaussian beam utilized in the final convolution is a parameter that has to be arbitrarily selected by the user, although estimates on maximum baseline coverage are usually adopted. As a consequence, CLEAN is a strongly supervised algorithm. It is key to the success of CLEAN that the astronomer performing the analysis builds in their perception of image structure in an interactive way by changing the CLEAN windows, gain, taper, and weighting during the imaging procedure. This makes CLEAN reconstructions often challenging to reproduce.

Multiple of these issues can be effectively solved by recently proposed variants of the standard CLEAN algorithm \citep[e.g.][]{Mueller2022b, Perracchione2023}. The classical way to increase resolution for CLEAN is by updating the weights, e.g. to uniform or even super-uniform weighting. Since most interferometric experiments have a higher density of short baselines rather than long ones, this leads to an overweighting of large scale structures at the sake of resolution. Uniform weighting, or any hybrid weighting in between \citep{Briggs1995}, addresses this fact by giving a larger weight to the long baselines, at the cost of overall structural sensitivity. 

The U-CLEAN algorithm recently proposed by \citet{Perracchione2023} combines the CLEAN method with an extrapolation/interpolation scheme that allows for a more automated CLEAN procedure. Specifically, the method utilizes the Fourier transform of the CLEAN components as a-priori information for performing a reliable interpolation of the real and the imaginary parts of the visibilities. This feature augmentation technique is particularly useful for image reconstruction from STIX data since, in that case, the interpolation task  suffers from the extreme sparsity of the $(u,v)$-coverage. Once the visibility interpolation step is completed, image reconstruction is performed by minimizing the reduced $\chi^2$ between the interpolated visibility surface and the Fourier transform of the image via a projected Landweber algorithm \citep{1996JOSAA..13.1516P}.

DoB-CLEAN is a recently proposed multiscalar CLEAN variant \citep{Mueller2022b} that models the image by the difference of elliptical Bessel functions (DoB-functions) that are fitted to the $(u,v)$-coverage. CLEAN is not used as a deconvolution algorithm, but as a feature finder algorithm. The cleaning of the image is performed by switching between the DoB-dictionary to a dictionary consisting of the difference of elliptical Gaussian functions. Both DoB-CLEAN and U-CLEAN waive the necessity for the final convolution of the CLEAN components with the CLEAN beam \citep{Mueller2022b, Perracchione2023} and, hence, the algorithms are not biased by the arbitrary choice of the CLEAN beam FWHM.

\subsection{Maximum Entropy}
Historically, Maximum Entropy Methods (MEM) were among the first algorithms that were proposed for imaging \citep[e.g.][]{Frieden1972, Ponsonby1973, Ables1974}. Despite its relative age, MEM was disfavored in practice in comparison to CLEAN. However, there are a number of active developments for MEM based algorithms \citep[e.g.][]{Massa2020, Mus2023}. Many of them reinterprete the MEM functional as one of many objectives in forward modeling frameworks \citep{Chael2016, Chael2018, Mueller2023, Mus2023b}. These algorithms are based on the constrained optimization framework~\citep[originally presented in ][]{Cornwell1985} or on the unconstrained optimization setting.

MEMs are regularization techniques that use a large image entropy as a prior information, i.e. among all the possible solutions that may fit the data the most simple (in the sense of information fields) is selected. If $I=I\left(x\right)$ and $G=G\left(x\right)$, the entropy is measured by the Kullback–Leibler divergence functional \citep{Kullback1951}:
\begin{align}
    R_{{\rm MEM}}(I) = -\int I \ln \left( \frac{I}{G} \right)\ \text{dx},
\end{align}
where $G$ is a possible prior image. Common priors include ``flat'' priors (i.e. a constant images rescaled in such a way that the sum of the pixels is equal to a priori estimate of the total flux), or Gaussian images corresponding to the size of the compact flux emission region of the image structure. 
To ensure that the proposed guess solution fits the data, the value of the reduced $\chi^2$ data-fitting metric is constrained. In particular, the problem presented in~\cite{Cornwell1985} is
\begin{equation*}\label{eq:Cons_MEM}
\tag{$\text{Cons\_MEM}$}
    \begin{aligned}
      & \underset{I}{\text{maximize}}
      & & -\displaystyle{\sum_i} I_i\log \dfrac{I_i}{G}~,\\
      & \text{subject to}
      & & \chi^{2}(V, \mathcal{F}I) - \Omega = 0~,\\
      & & & F^{\text{mod}}-\displaystyle{\sum_i}I_i=0~,
    \end{aligned}
\end{equation*}
where $V$ is the array of visibilities, $\mathcal{F}$ is the forward operator (undersampled forward operator), $F^{\text{mod}}$ is the model flux and $\Omega$ a noise estimation. To satisfy non-negativity of the solution, we model the image by a lognormal transform, a strategy that has become popular in Bayesian algorithms \citep{Arras2019, Arras2021}.

From the optimization view-point, the main disadvantage of the problem \eqref{eq:Cons_MEM} is its non-convexity, given the quadratic equality constraint defined by the reduced $\chi^2$. Therefore, \citet{Massa2020} defined a new version of the maximum entropy problem, named MEM\_GE, in which the objective function is a weighted sum of the reduced $\chi^2$ and of the negative entropy functional:
\begin{equation*}\label{eq:MEM_GE}
\tag{$\text{MEM\_GE}$}
\begin{aligned}
& \underset{I}{\text{minimize}}
& & \chi^{2}(V, \mathcal{F}I) +\lambda \displaystyle{\sum_i} I_i\log \dfrac{I_i}{G},\\
& \text{subject to}
& &F^{\text{mod}}-\displaystyle{\sum_i}I_i=0 ~, \\
& & &I_i \geq 0 \quad \forall i
\end{aligned}
\end{equation*}
where $\lambda$ is the regularization parameter balancing the trade-off between data-fitting and regularization. Further, in the MEM\_GE implementation $G$ is chosen as a ``flat'' prior. The optimization problem \eqref{eq:MEM_GE} is convex and can be solved by standard optimization techniques. Specifically, \citet{Massa2020} adopted an accelerated forward-backward splitting \citep{beck2009fast,combettes:hal-00643807}. We note that the MEM\_GE approach is similar to Regularized Maximum Likelihood approaches (cf. Sect. \ref{sec:RML}). However, the latter often combines several regularization terms and, therefore, makes a single description by a proximal point minimization method more challenging. Therefore, the minimization techniques that are typically adopted in RML approaches are gradient descent, conjugate gradient (CG)~\citep{CG} or limited-memory BFGS (L-BFG-S)~\cite{LBFGS}. In this manuscript, among others, we compare the latter strategy for entropy objectives (RML$\_$MEM) with the MEM$\_$GE approach. A further difference between MEM\_GE and RML approaches is in the assumed prior distribution (``flat'' prior versus Gaussian prior) and in the automatic stepsize and balancing update developed for MEM$\_$GE \citep{Massa2020}.

\subsection{Regularized Maximum Likelihood}\label{sec:RML}
Regularized Maximum Likelihood (RML) methods approach the imaging problem by balancing data fidelity terms and regularization terms, i.e. we solve an optimization problem of the form:
\begin{align}
    \hat{I} \in \argmin_I \left\{ \sum_i \alpha_i S_i(V, \mathcal{F}I) + \sum_j \beta_j R_j(I) \right\}, \label{eq: rml}
\end{align}
where $V$ are the observed visibilities, $\mathcal{F}$ the forward operator (undersampled Fourier transform), $S_i$ the data fidelity terms and $R_j$ the regularization terms. $\alpha_i, \beta_j \in \mathbb{R}_+$ are regularization parameters that control the balancing between the different terms. 
The data fidelity terms measure the fidelity of the guess solution $I$ and the regularization term measures the feasibility of the solution according to the perception of the image structure. Usual data fidelity terms are reduced $\chi^2$-metrics between the observed visibilities and the predicted visibilities. For VLBI, the use of calibration independent closure quantities has become more common \citep{Chael2018, Mueller2022, Mueller2023}, but for this work we use the reduced $\chi^2$-metric to the visibilities. The regularization terms encode various prior assumptions on the feasibility of the image, e.g. simplicity ($l^2$-norm), sparsity ($l^1$-norm), smoothness (total variation, for the remainder of this manuscript abbreviated by TV, and total squared variation), maximal entropy (Kullback-Leibler divergence) or a total flux constraint. For a full list of regularization terms that were used in VLBI we refer the interested reader to the discussions in \citet{eht2019d}. While RML methods are expected to produce excellent, super-resolving images \citep{ngehtchallenge, eht2019d}, they depend strongly on the correct regularization parameter selection $\alpha_i, \beta_j$. For the sake of simplicity we focus on entropy, sparsity and total variation terms for this manuscript and choose a representative, but not necessarily ideal parameter combination. Particularly, we test following approaches:
\begin{align}
    &\hat{I} \in \argmin_I \left\{ \chi^2(V, \mathcal{F}I) + \alpha \norm{I}_{l1}\right\}   \tag{$\text{l1}$} \\
    &\hat{I} \in \argmin_I \left\{ \chi^2(V, \mathcal{F}I) + \alpha |I|_{TV}\right\}   \tag{$\text{TV}$}\\
    &\hat{I} \in \argmin_I \left\{ \chi^2(V, \mathcal{F}I) + \alpha \int I \ln\frac{I}{G}\ \text{dx}\right\}   \tag{$\text{MEM}$}\\
    &\hat{I} \in \argmin_I \left\{ \chi^2(V, \mathcal{F}I) + \alpha \int I \ln\frac{I}{G}\ \text{dx} + \beta \norm{I}_{l1}  \right\} \tag{$\text{MEM-l1}$}\\
    &\hat{I} \in \argmin_I \left\{ \chi^2(V, \mathcal{F}I) + \alpha \int I \ln\frac{I}{G}\ \text{dx} + \beta |I|_{TV} \right\} \tag{$\text{MEM-TV}$}\\
    &\hat{I} \in \argmin_I \left\{ \chi^2(V, \mathcal{F}I) + \alpha |I|_{TV} + \beta \norm{I}_{l1} \right\} \tag{$\text{TV-l1}$}\\
    &\hat{I} \in \argmin_I \left\{ \chi^2(V, \mathcal{F}I) + \alpha \int I \ln\frac{I}{G}\ \text{dx} +  \beta |I|_{TV} + \gamma \norm{I}_{l1} \right\} \tag{$\text{MEM-TV-l1}$},\\
\end{align}
with respective regularization parameters $\alpha, \beta, \gamma$. In practice, one needs to ensure that the pixels are not negative. This is handled by the lognormal transform, i.e. apply the change of variables $I \mapsto \exp I$, before evaluating the Fourier transform \citep{Chael2018, Arras2022}. The prior for the entropy functional $G$ is chosen as a Gaussian whose full width at half maximum (FWHM) is consistent with the size of the compact emission region. The size of the compact emission region may be evaluated in practice from multiwavelength studies of the source \citep{eht2019d}.

\subsection{Compressive Sensing}
Compressive Sensing (CS) aims at sparsely representing the image in a specific basis. Typically wavelets have been used for this task \citep[e.g. see ][for application within radio interferometry]{Candes2006, Donoho2006, Starck2006, Starck2015, Mertens2015, Line2020}. The sample of basis functions is called a dictionary $\Gamma$, the single basis functions are called atoms. In the following we will choose $\Gamma$ as a wavelet dictionary. The idea behind the CS technique is to represent the image as a linear combination of the atoms, i.e. $I = \Gamma \mathcal{I}$, and to recover the array of coefficients $\mathcal{I}$ by solving:
\begin{align}
    \hat{\mathcal{I}} \in \argmin_{\mathcal{I}} \left\{ S(V, \mathcal{F} \Gamma \mathcal{I}) + \beta \norm{\mathcal{I}}_{l1} \right\}.
\end{align}
Such algorithms and many variants have been studied in radio interferometry for a long time \citep{Weir1992, Bontekoe1994, Starck1994, Pantin1996, Starck2001, Maisinger2004, Li2011, Carrillo2012, Carrillo2014, Garsden2015, Girard2015, Onose2016, Onose2017, Cai2018a, Cai2018b,  Pratley2018, Mueller2022, Mueller2022b}. For the comparison that was carried out for this work, we consider the \textit{DoG-HiT} algorithm that was recently proposed by \citet{Mueller2022, Mueller2023b}. \textit{DoG-HiT}, standing for Difference-of-Gaussian Hard Iterative Thresholding, utilizes difference of Gaussian wavelets functions. For \textit{DoG-HiT} the basis functions are fitted to the $(u,v)$-coverage, hence optimally separating measured and non-measured Fourier coefficients. The minimization problem is solved with an iterative forward-backward splitting technique. \textit{DoG-HiT} was proven to recover images of comparable quality as RML methods \citep{Mueller2022, ngehtchallenge}, although the optimization landscape is much simpler. With only one free regularization parameter, \textit{DoG-HiT} is a substantial step towards an unsupervised imaging algorithm without the need of large parameter surveys. \textit{DoG-HiT} was originally proposed for closure-only imaging; for this comparison, we fit the visibilities directly instead.

\subsection{Multiobjective Imaging}
Multiobjective Optimization (MOEA/D) is a recently proposed \citep{Mueller2023, Mus2023b} imaging algorithm for VLBI. It mimics the formulation of the imaging problem in the RML framework, i.e. with a set of data fidelity terms and regularization terms. However, instead of solving a weighted sum of these terms as in Eq. \eqref{eq: rml}, we solve a multiobjective problem consisting of all the regularizers and data terms as single objectives \citep[for more details we refer to][]{Mueller2023}. A solution to the multiobjective problem is called Pareto optimal if the further optimization along one objective automatically has to worsen another one. The set of all Pareto optimal solutions is called the Pareto front. MOEA/D recovers the Pareto front. Since several regularizers introduce conflicting assumptions (i.e. sparsity in comparison to smoothness) the Pareto front divides into a number of disjunct clusters \citep{Mueller2023}, everyone describing a locally optimal mode of the multimodal reconstruction problem. In this spirit, it is not the goal of MOEA/D to recover a single image, but to recover a full hypersurface of image structures. This is most easily realized with the help of evolutionary algorithms \citep{Zhang2008, Li2009}. In order to select the objectively best cluster of representative solutions, we apply the accumulation point selection criterion proposed in \citet{Mueller2023}: we select the cluster of images that has the largest number of members in the final population.

\subsection{Bayesian Imaging}
Bayesian reconstruction methods are intensively studied in radio-interferometry \citep[e.g. see recently][]{Junklewitz2016, Cai2018a, Cai2018b, Arras2019, Broderick2020, Broderick2020b, Arras2021, Arras2022, Tiede2022, Roth2023, Kim2024}. Bayesian imaging methods calculate the posterior sky distribution $\mathcal{P}(I|V)$ from the prior distribution $\mathcal{P}(I)$ and visibility data $V$ by Bayes' theorem:
\begin{align} \label{eq:Bayes' thm}
\mathcal{P}(I|V) = \frac{\mathcal{P}(V|I)\, \mathcal{P}(I)} {\mathcal{P}(V)},
\end{align}
where $\mathcal{P}(V|I)$ is the likelihood distribution, $\mathcal{P}(V)$ is the evidence. The prior distribution contains the prior knowledge of the source $I$, such as positivity and smoothness constraints, and the likelihood represents our knowledge of the measurement process.

In Bayesian imaging, instead of obtaining an image, samples of possible images are reconstructed; therefore, it allows us to estimate the mean and standard deviation from the samples. Bayesian imaging has a distinctive feature: the uncertainty information in the visibility data $V$ domain can be propagated in other domains. As a consequence, the reliability of reconstructed parameters, such as image $I$ and antenna gain, can be quantified by the uncertainty estimation.

Bayes' theorem can be written by the negative log-probability, also called the energy or Hamiltonian $\mathcal{H}$ \citep{Ensslin2019}:
\begin{align} \label{eq:posterior Hamiltonian}
    \mathcal{H}(I|V) \equiv - \ln(\mathcal{P}(I|V)) = \mathcal{H}(V|I) + \mathcal{H}(I) - \mathcal{H}(V).
\end{align}
The maximum a-posteori sky posterior distribution $I$ is determined by minimizing an objective function containing the likelihood and prior:
\begin{align} \label{Bayesian_objective_ftn}
    \hat{I}_{\mathrm{MAP}} \in \argmin_{I} \left\{ 
    \mathcal{H}(V|I) + \mathcal{H}(I)
    \right\},
\end{align}
where $\mathcal{H}(V|I)$ is the likelihood Hamiltonian and $\mathcal{H}(I)$ is the prior Hamiltonian. Note that the evidence term is ignored in the objective functional since it does not depend on the sky brightness distribution $I$.

From the posterior sky brightness distribution $\mathcal{P}(I|V)$, we can calculate the posterior mean:
\begin{align}
\hat{I} = \int I \, \mathcal{P}(I|V) dI.
\end{align}

Note that the reconstructed image in this work is the posterior sky mean $I$. Furthermore, the standard deviation of sky $I$ can be calculated from the posterior distribution $\mathcal{P}(I|V)$ analogously. The standard deviation allows us to quantify the reliability of reconstructed results.

Bayesian inference requires substantial computational resources in order to obtain the posterior probability distribution instead of a scalar value for each parameter. For instance, sampling by a full-dimensional Markov-Chain Monte-Carlo (MCMC) procedure, or evalutaing the high-dimensional integrals of the mean directly, takes often too much time \citep{Cai2018a, Cai2018b}. 

In this work, images are reconstructed by means of the Bayesian imaging algorithm \textit{resolve}. In order to perform high-dimensional image reconstruction, Variational Inference (VI) methods \citep{Knollmueller2019, Frank2021} are used in \textit{resolve}. In Variational Inference method, the Kullback-Leibler divergence \citep{Kullback1951} is minimized in order to find approximated posterior distribution as closely as the true posterior distribution. Although the uncertainties tend to be underestimated in the VI method, it enables us to estimate high-dimensional posterior distribution. In conclusion, it strikes a balance between the performance of the algorithm and the statistical integrity.

Bayesian imaging shares some similarities with RML approaches, in the sense that the RML methods can be interpreted as the maximum a posteriori (MAP) estimation in Bayesian statistics. The negative log-likelihood $\mathcal{H}(V|I)$ in Eq. \ref{Bayesian_objective_ftn} is equivalent to the data fidelity term and the negative log-prior $\mathcal{H}(I)$ can be interpreted as the Bayesian equivalent to the regularizer in RML methods \citep{Kim2024}. For instance, the correlation structure between image pixels can be inferred by Gaussian process with non-parametric kernel in \textit{resolve}\citep{Arras2021}. The unknown correlation structure can be inferred by the data, which plays a similar role as a smoothness prior in RML methods.

\section{Challenge} \label{sec: challenge}

\subsection{Synthetic Data Sets}
To test this variety of algorithmic frameworks we compare our results on a set of synthetic data. These include typical image structures that could be expected for the STIX and the EHT instruments. Particularly, we study the following synthetic data sets.

STIX synthetic data sets replicate a solar flare. To mimic spatial features that may be expected for observations with STIX, we simulate a double Gaussian structure and a loop shape \citep[see][for more details on the definition of the considered parametric shapes]{Volpara2022}. In particular, the double Gaussian structure represents typical non-thermal X-ray sources at the flare footpoints, while the the loop shape represents a thermal source at the top of the flare loop.
Further, the two Gaussian sources have different size and different flux: the left source has a FWHM of 15 arcsec and flux equal to 66\% of the total flux, while the right source has a FWHM of 10 arcsec and flux equal to 33\% of the total flux. We generated synthetic STIX visibilities corresponding to these configurations by computing their Fourier transform in the frequencies sampled by STIX. For the experiments performed in this paper, we considered only the visibilities associated with the 24 sub-collimators with coarsest angular resolution. Indeed, the remaining six sub-collimators have not yet been considered for image reconstruction since their calibration is still under investigation \citep{massa2022hard}.

We add uncorrelated Gaussian noise to every visibility. To study the effect of the noise-level on the final reconstruction, we prepared three different data sets with varying noise-levels. We note that statistical errors affecting the STIX visibilities are due to the Poisson noise of the photon counts recorded by the STIX detectors. We simulated three different levels of data statistics corresponding to a number of counts per detector equal to \(\sim\)500, \(\sim\)2500, \(\sim\)5000. These data-statistics will be referred to as low S/N, medium S/N and high S/N for the remainder of the manuscript. Then, we added Gaussian noise to the visibilities with a standard deviation that is derived from the Poisson noise affecting the count measurements, resulting in an effective median S/N of $\approx$ 5, 12, and 16 respectively. Due to the small number of visibility points for STIX, the reconstruction is sensitive to the random seed of the random noise distribution. To get a statistical assessment on the reconstruction quality, we recover the images from ten different realizations of the STIX data with a varying seed for the random noise distribution and average the results. In Fig. \ref{fig: double_gauss_medium_stat}, \ref{fig: loop_medium_stat} we show the average over the ten reconstructions.

Since the similarity between STIX and VLBI imaging is greatest (in terms of accessible dynamic range, spatial scales and degree of undersampling) when the VLBI $(u,v)$-coverage is sparsest, and this is the data regime that saw a rapid development of novel methods we scrutinize in this work \citep[e.g.][]{Chael2016, Chael2018, Akiyama2017, Akiyama2017b, Arras2022, Mueller2022, Mueller2022b, Mueller2023, Mus2023}, we focus here on geometric models that mimick EHT observations. In fact, we study a crescent geometric model. This model is a simple geometric approximation to the first image of the shadow of the supermassive black hole in M87 presented by \citet{eht2019a}. It was particularly used for the verification of the imaging strategies for the analysis of M87 \citep{eht2019d}, as well as SgrA* \citep{eht2022c}. We add thermal noise consistent with the system temperatures reported in \citet{eht2019d}, and we assume that the phase and amplitude calibration is known.

The next generation EHT (ngEHT) is a planned extension of the EHT that is supposed to deliver transversely resolved images of the black hole shadow \citep{Doeleman2019, Johnson2023}. In order to test the algorithmic needs of this future frontline VLBI project, we consider also synthetic data inspired by the first ngEHT analysis challenge presented by \citet{ngehtchallenge}. We use a General Relativistic Magneto Hydrodynamic (GRMHD) simulation of the supermassive black hole M87 \citep{Mizuno2021, Fromm2022}. We generated a synthetic set of data corresponding to the proposed ngEHT configuration, which consists of the current EHT antennas and ten additional proposed antennas \citep[see e.g. for more details][]{Raymond2021, ngehtchallenge}. We add thermal noise, and assume phase and amplitude calibration. For VLBI, due to the larger number of visibilities with uncorrelated Gaussian noise contribution, we do not need to study several realizations of the noise contribution.

STIX and VLBI (and particularly the EHT) utilize various astronomical conventions, data formats, software packages, and various programming languages for the respective data analysis. To transfer the STIX data sets into a VLBI framework, we created a virtual VLBI snapshot observation that has exactly the same $(u,v)$-coverage as STIX.  
Vice versa, we extract the $(u,v)$-coverage, visibilities, and noise ratios from the VLBI observations, and save the related data arrays and matrices in a readable format for STIX.

\subsection{Comparison Metrics}
We compare the reconstruction results to the ground truth images using three different metrics, inspired by the metrics that were used in the recent imaging challenge presented by \citet{ngehtchallenge}. First, and most importantly, we compute the match between the ground truth images and the recovered images by means of the cross correlation:
% \begin{align}
% \rho_{NX} = \frac{1}{N} \sum_{i=1}^N \frac{\left( X_i- \langle X \rangle\right)\left( Y_i- \langle Y \rangle \right)}{\sigma_X \sigma_Y} ~,
% \end{align}
\begin{align}
\rho = \frac{1}{N} \sum_{j=1}^N \frac{\left( I^{\rm GT}_j- \langle I^{\rm GT} \rangle\right)\left( I^{\rm R}_j- \langle I^{\rm R} \rangle \right)}{\sigma_{I^{\rm GT}} \sigma_{I^{\rm R}}} ~,
\end{align}
where $I^{\rm GT}$ and $I^{\rm R}$ represent the ground truth and the recovered image, respectively, $\langle ~\cdot~ \rangle$ denotes the mean of the image pixel values and $\sigma$ their standard deviation.
This metric has been used in the past to determine the precision of VLBI algorithms in the framework of the EHT \citep{eht2019d, ngehtchallenge}. 

Moreover, we calculate the effective resolution of a reconstruction. Since the reconstructions are typically more blurred than the ground truth image, we determine the effective resolution by the following strategy. We blur the ground truth image gradually with a circular Gaussian beam and compute $\rho$ between the recovered image and the blurred ground truth image. We do this for a predefined set of blurring kernels, and select the one that maximizes $\rho$.

Finally, we determine the dynamic range with a strategy inspired by the proxy proposed in \citet{Bustamante2023}. To have an approximation for the dynamic range that is independent of the resolving power of an algorithm, we first blur every recovered image to the nominal resolution $\theta_{\rm nom}$, i.e. the width of the clean beam. We do this in a similar way that we applied to estimate the image resolution. We gradually blur the recovered images with a blurring kernel and calculate the cross correlation $\rho$ between the blurred recovered image and the ground truth image, that was blurred with the nominal resolution. We finally blur the recovered images with the beam (Gaussian with width $\theta_{\rm res}$) that maximizes the correlation to the true image at nominal resolution. Then we compute the proxy of the dynamic range in the image: 
\begin{align}
\mathcal{D} = \frac{\max \left( I^{\rm GT} * \mathcal{G}_{\theta_{\rm nom}} \right)}{|I^{\rm GT} * \mathcal{G}_{\theta_{\rm nom}} - I^{R} * \mathcal{G}_{\theta_{\rm res}}|} ~,
\end{align}
where $\mathcal{G}_{\theta}$ denotes a circular Gaussian with standard deviation equal to $\theta$.
We report the $q$-th quantile of $\mathcal{D}$, where we choose $q=0.1$ in consistency with \citet{ngehtchallenge, Bustamante2023}:
\begin{align}
D_{0.1} = \mathrm{quantile}(\mathcal{D}, q)|_{q=0.1} ~.
\end{align}

The numerical complexity of an algorithm is an additional important benchmark. Note that imaging algorithms proposed for VLBI (and radio interferometry in general) were historically proposed for bigger arrays with a larger number of visibilities, rendering them as relatively fast in a STIX setting, comparable in speed to the algorithms that were proposed for STIX directly. The running time of the single algorithms (once the hyperparameters are fixed) is not a concern and allows for nearly real time image analysis due to the small number of visibilities that were adapted for these examples. In this manuscript, we do not touch on the question how well the various algorithms scale to bigger data sizes as are common for denser radio interferometers in general. With a relatively small computational effort to evaluate the model visibilities, the time for user-interaction and finetuning of the software specific hyperparameters becomes more relevant to the overall time that is consumed for the reconstructions.

Some algorithms, as for example \textit{DoG-HiT}, MOEA/D, Cons\_MEM and MEM\_GE allow for an automatized imaging with minimal interaction, other algorithms depend on the fine-tuning of a varying number of hyperparameters (e.g., RML and Bayesian algorithms), or manual interaction (CLEAN). Given these considerations, it is challenging to assign a quantitative metric on the numerical and application complexity of the algorithms, since the actual time needed to set up and run the algorithms is prone to external factors such as the quality of the data set or the experience of the user. We therefore opt for a qualitative comparison for the computational resources needed and report on our experiences with the various data sets and imaging algorithms.

\subsection{Results}

\begin{figure*}
    \centering
    \begin{subfigure}[b]{1\textwidth}
        \includegraphics[width=\textwidth]{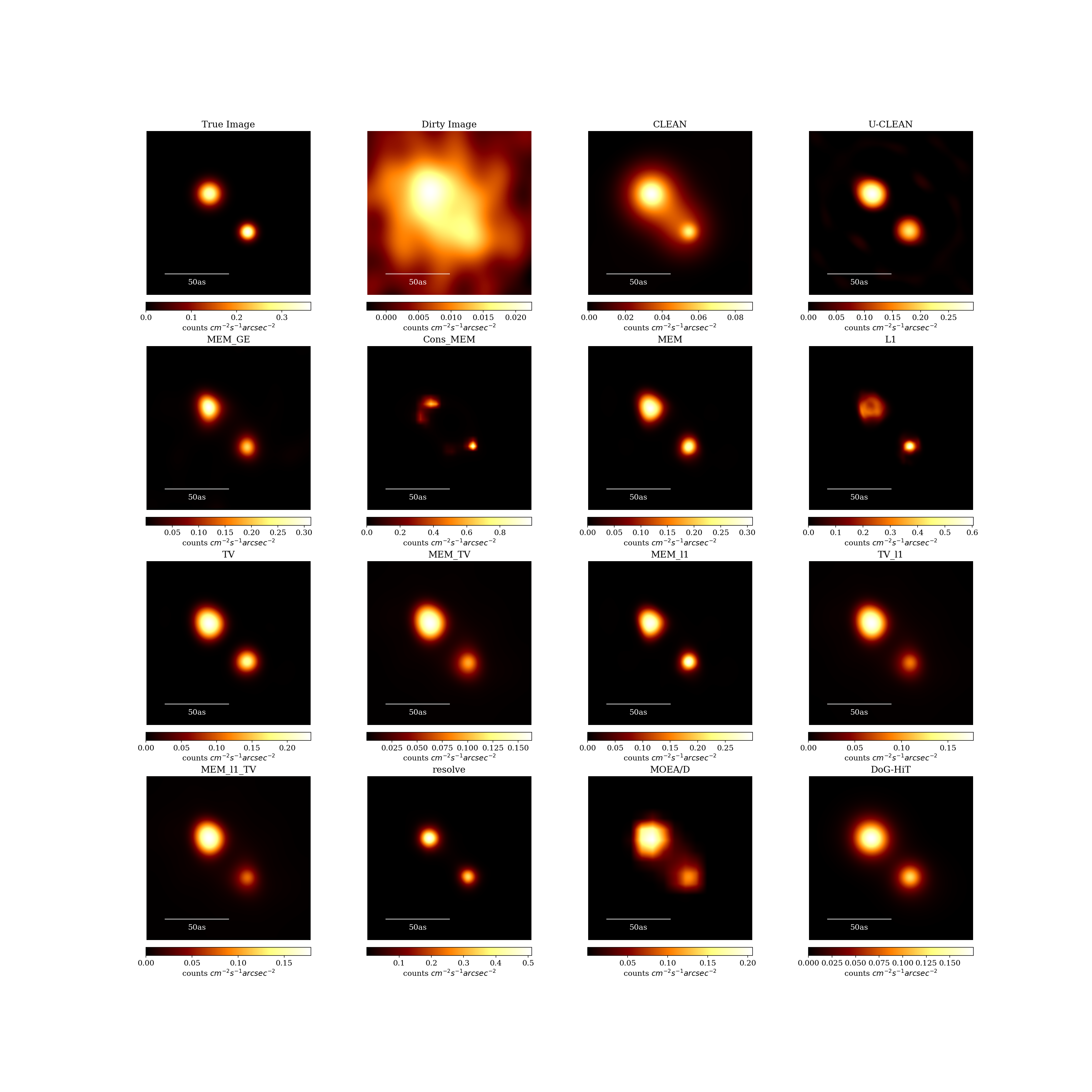}
        \caption{Reconstructions}
    \end{subfigure} \\
    \begin{subfigure}[b]{0.3\textwidth}
        \includegraphics[width=\textwidth]{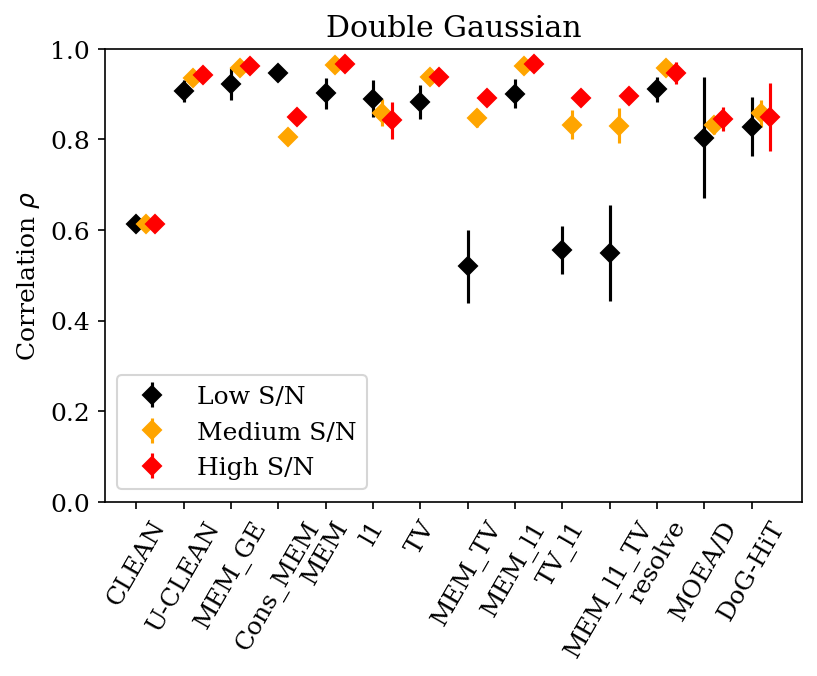}
        \caption{Correlation}
    \end{subfigure}
    \begin{subfigure}[b]{0.3\textwidth}
        \includegraphics[width=\textwidth]{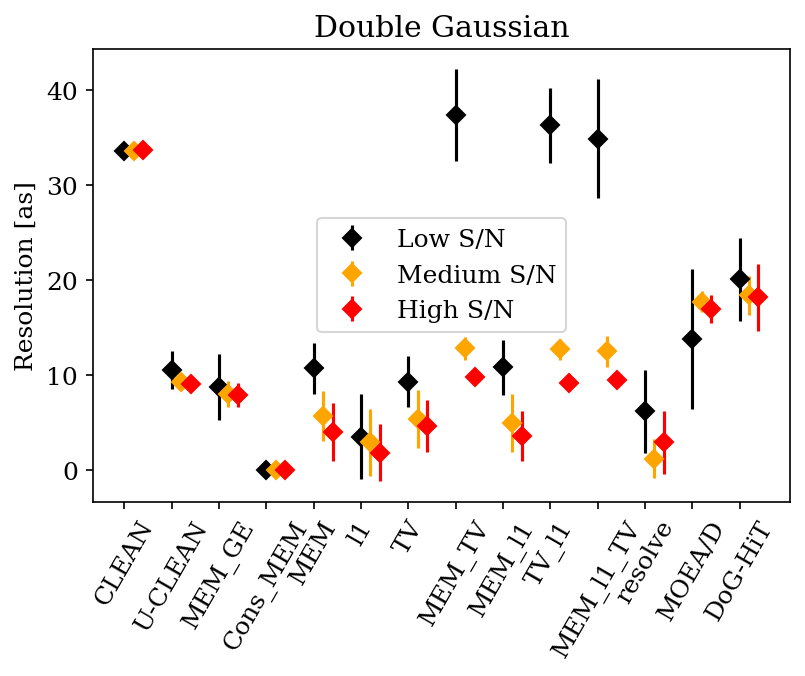}
        \caption{Resolution}
    \end{subfigure}
    \begin{subfigure}[b]{0.3\textwidth}
     \includegraphics[width=\textwidth]{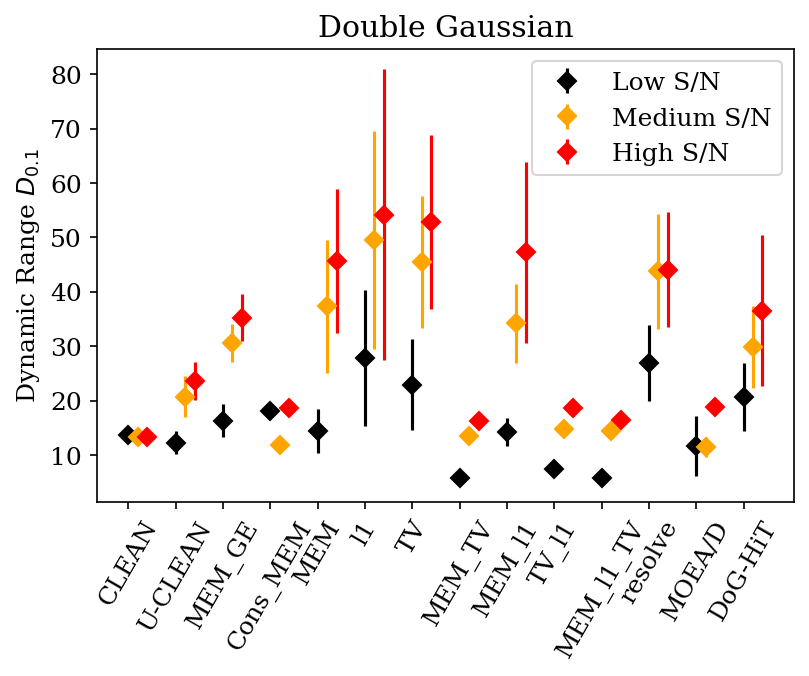}
     \caption{Dynamic Range}
    \end{subfigure}
    \caption{Reconstruction results and scoring of the various imaging algorithms on the double gaussian source model for STIX. In panel (a) we show the reconstruction results for medium noise levels. In the lower panels we compare the scoring of the reconstructions across various methods and noise-levels. Cons\_MEM is over-resolving the source.}
    \label{fig: double_gauss_medium_stat}
\end{figure*}

\begin{figure*}
    \centering
    \begin{subfigure}[b]{1\textwidth}
        \includegraphics[width=\textwidth]{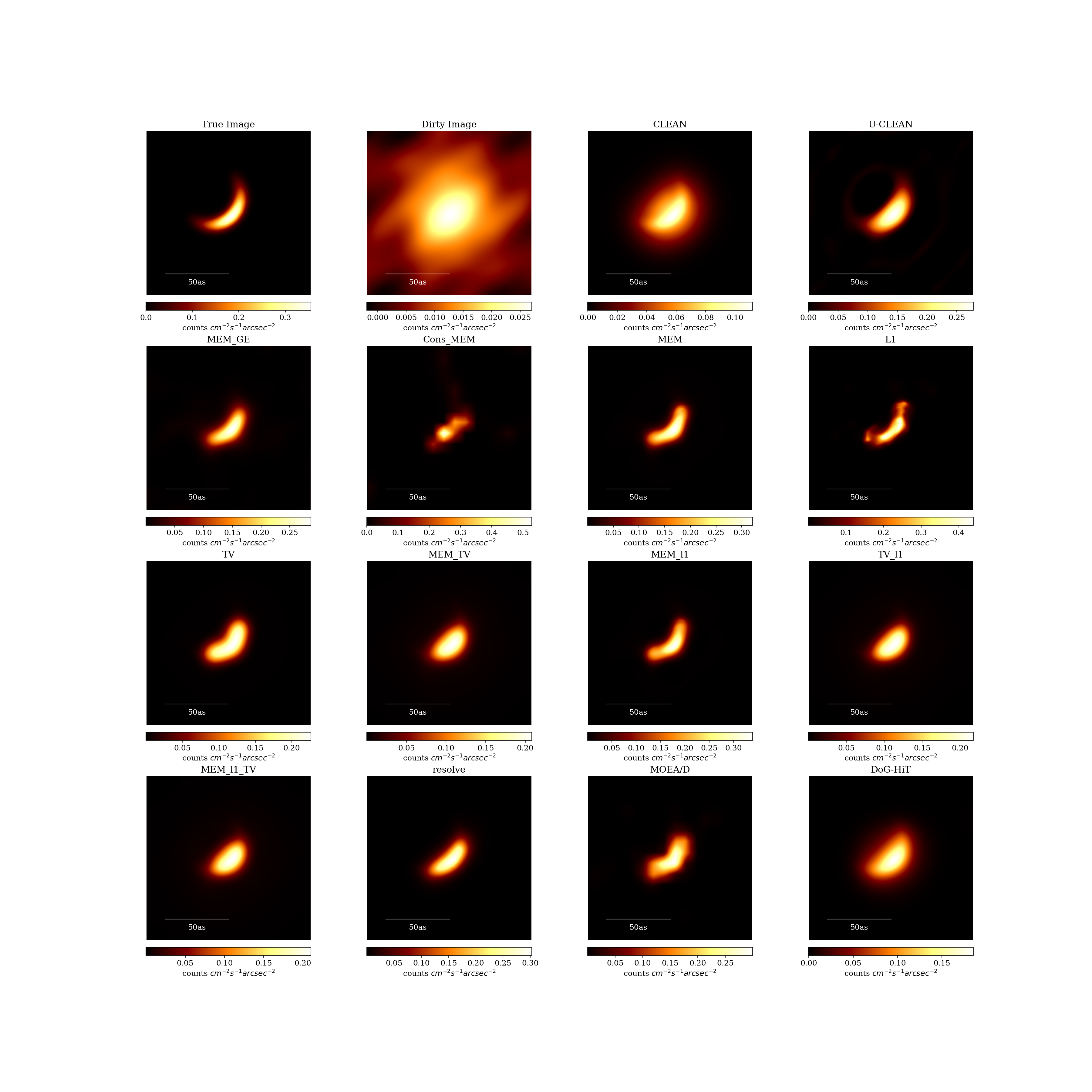}
        \caption{Reconstructions}
    \end{subfigure} \\
    \begin{subfigure}[b]{0.3\textwidth}
        \includegraphics[width=\textwidth]{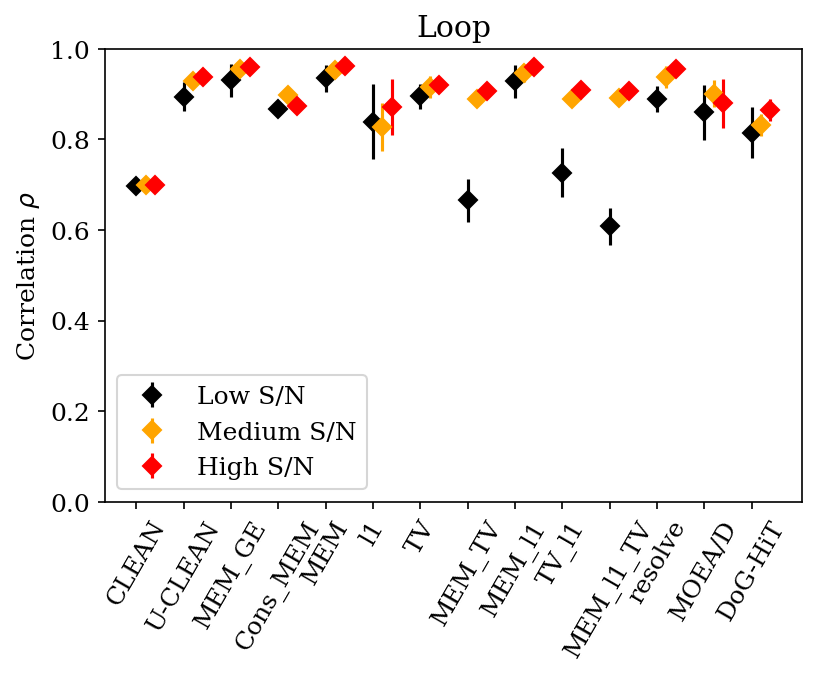}
        \caption{Correlation}
    \end{subfigure}
    \begin{subfigure}[b]{0.3\textwidth}
        \includegraphics[width=\textwidth]{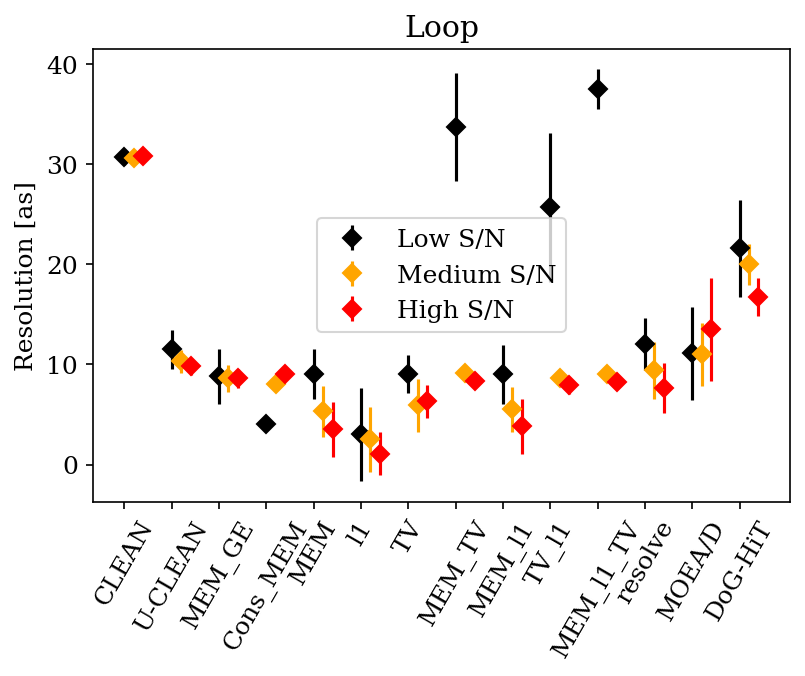}
        \caption{Resolution}
    \end{subfigure}
    \begin{subfigure}[b]{0.3\textwidth}
     \includegraphics[width=\textwidth]{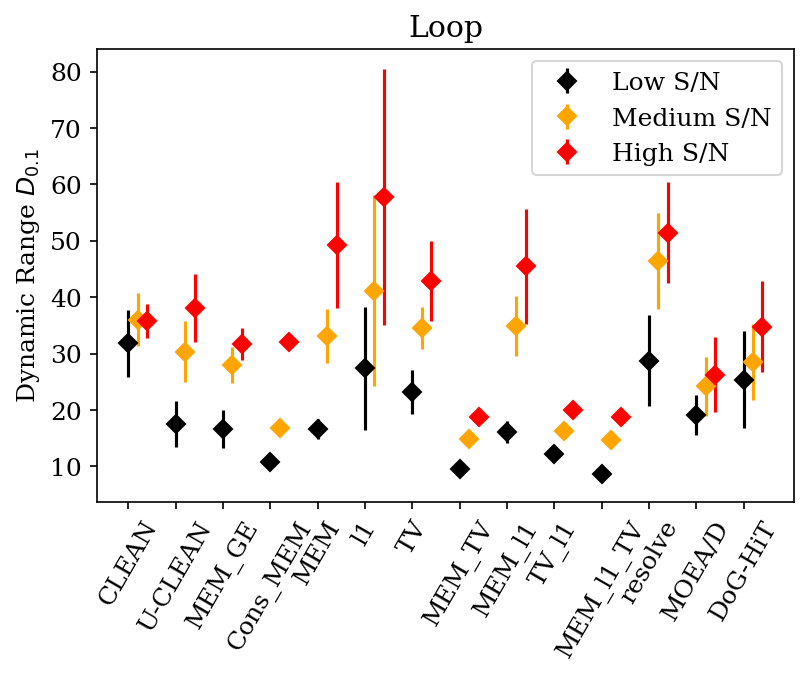}
     \caption{Dynamic Range}
    \end{subfigure}
    \caption{Same as Fig. \ref{fig: double_gauss_medium_stat}, but for the synthetic loop configuration.}
    \label{fig: loop_medium_stat}
\end{figure*}

\begin{figure*}
    \centering
    \begin{subfigure}[b]{1\textwidth}
        \includegraphics[width=\textwidth]{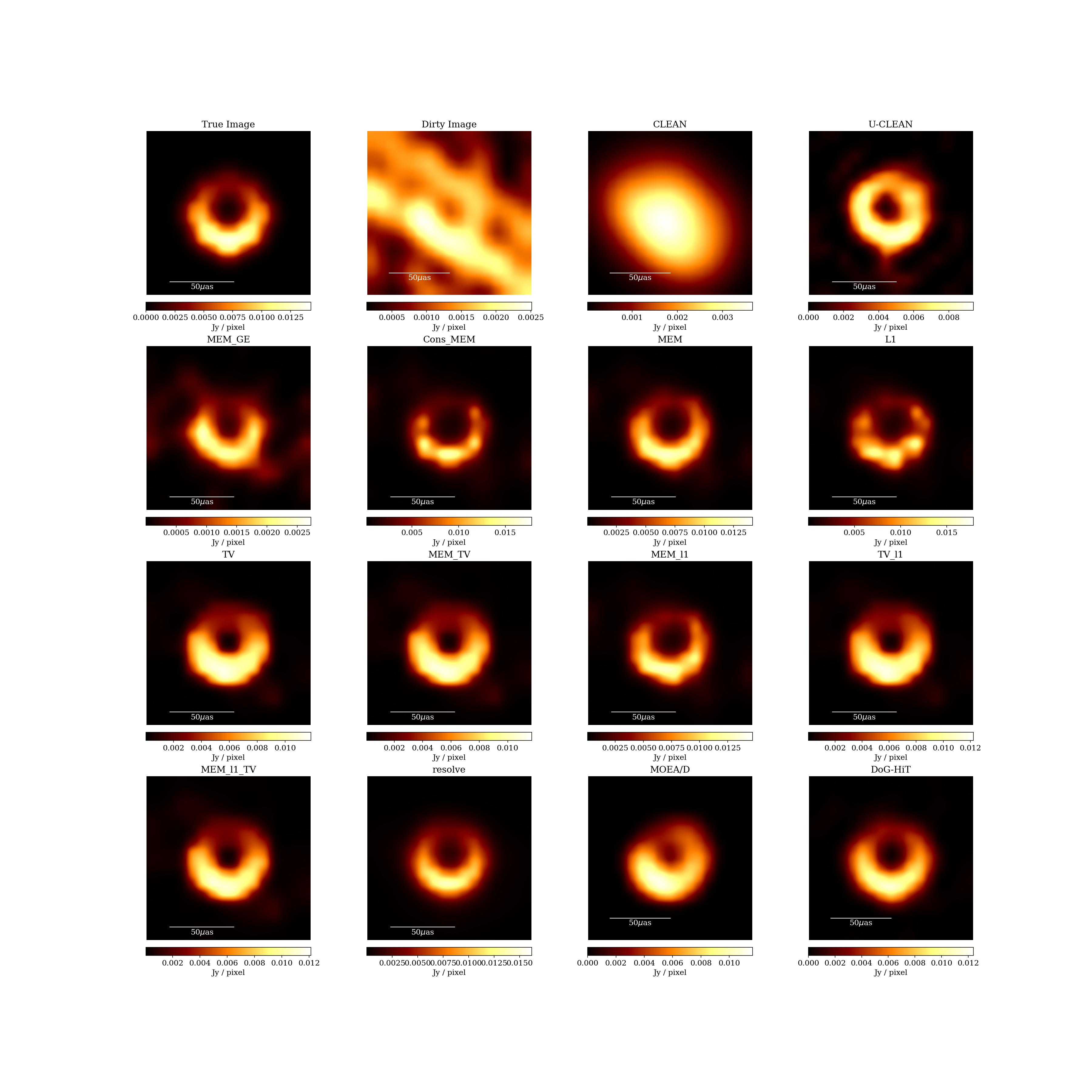}
        \caption{Reconstructions}
    \end{subfigure} \\
    \begin{subfigure}[b]{0.3\textwidth}
        \includegraphics[width=\textwidth]{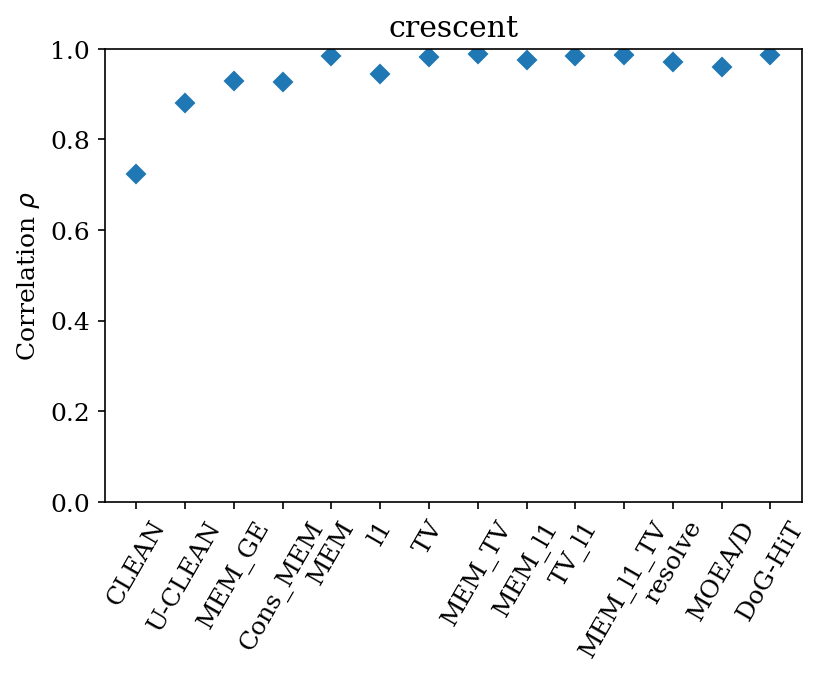}
        \caption{Correlation}
    \end{subfigure}
    \begin{subfigure}[b]{0.3\textwidth}
        \includegraphics[width=\textwidth]{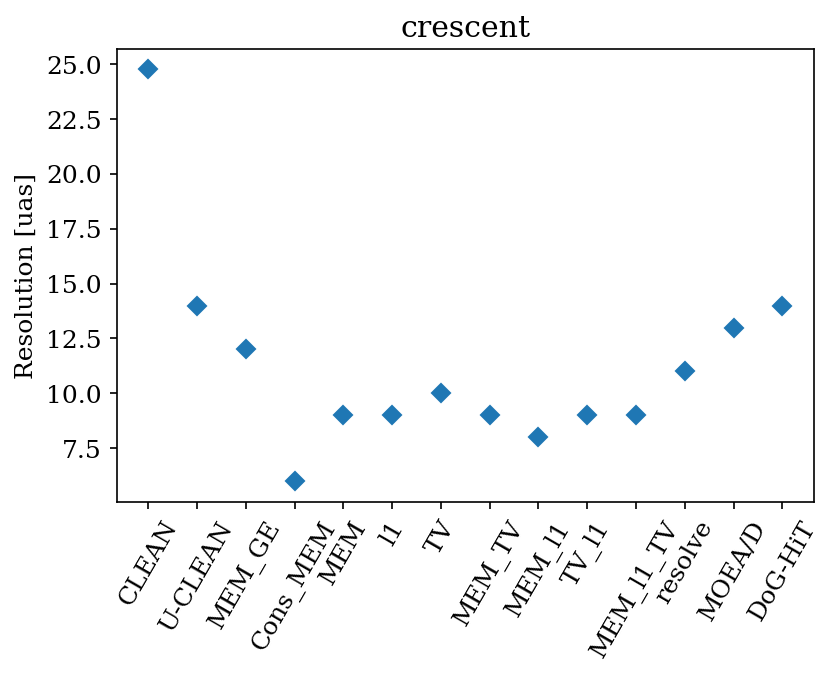}
        \caption{Resolution}
    \end{subfigure}
    \begin{subfigure}[b]{0.3\textwidth}
     \includegraphics[width=\textwidth]{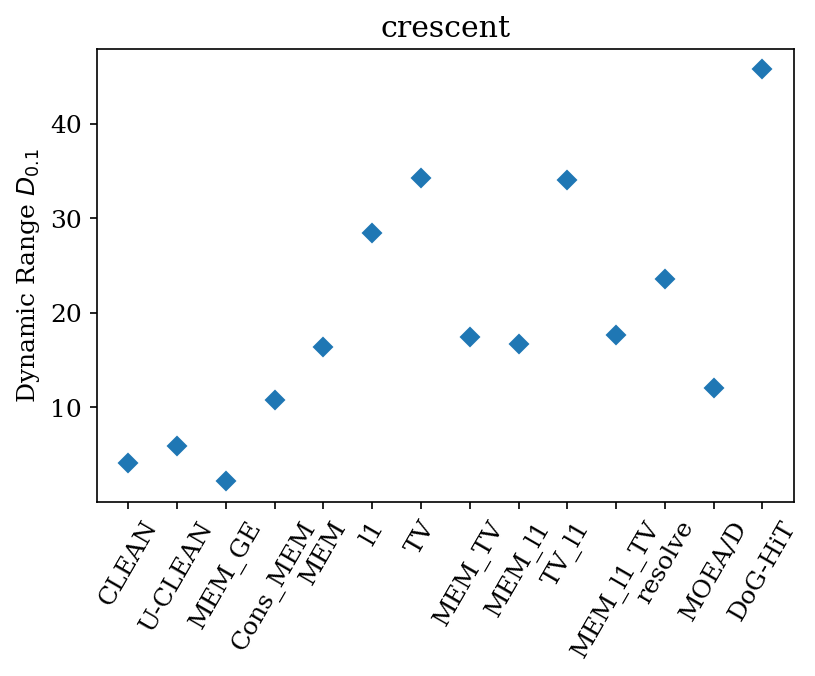}
     \caption{Dynamic Range}
    \end{subfigure}
    \caption{Reconstruction results and scoring of the various imaging algorithms on the crescent source model observed with the EHT configuration. Note that resolution values are close to zeros for a few algorithms (Cons\_MEM, MEM, l1 and MEM\_l1) since they over-resolve the source, i.e. they recover structures that are finer than those in the ground truth image due to too much enhancing of the regularizer.}
    \label{fig: crescent}
\end{figure*}

\begin{figure*}
    \centering
    \begin{subfigure}[b]{1\textwidth}
        \includegraphics[width=\textwidth]{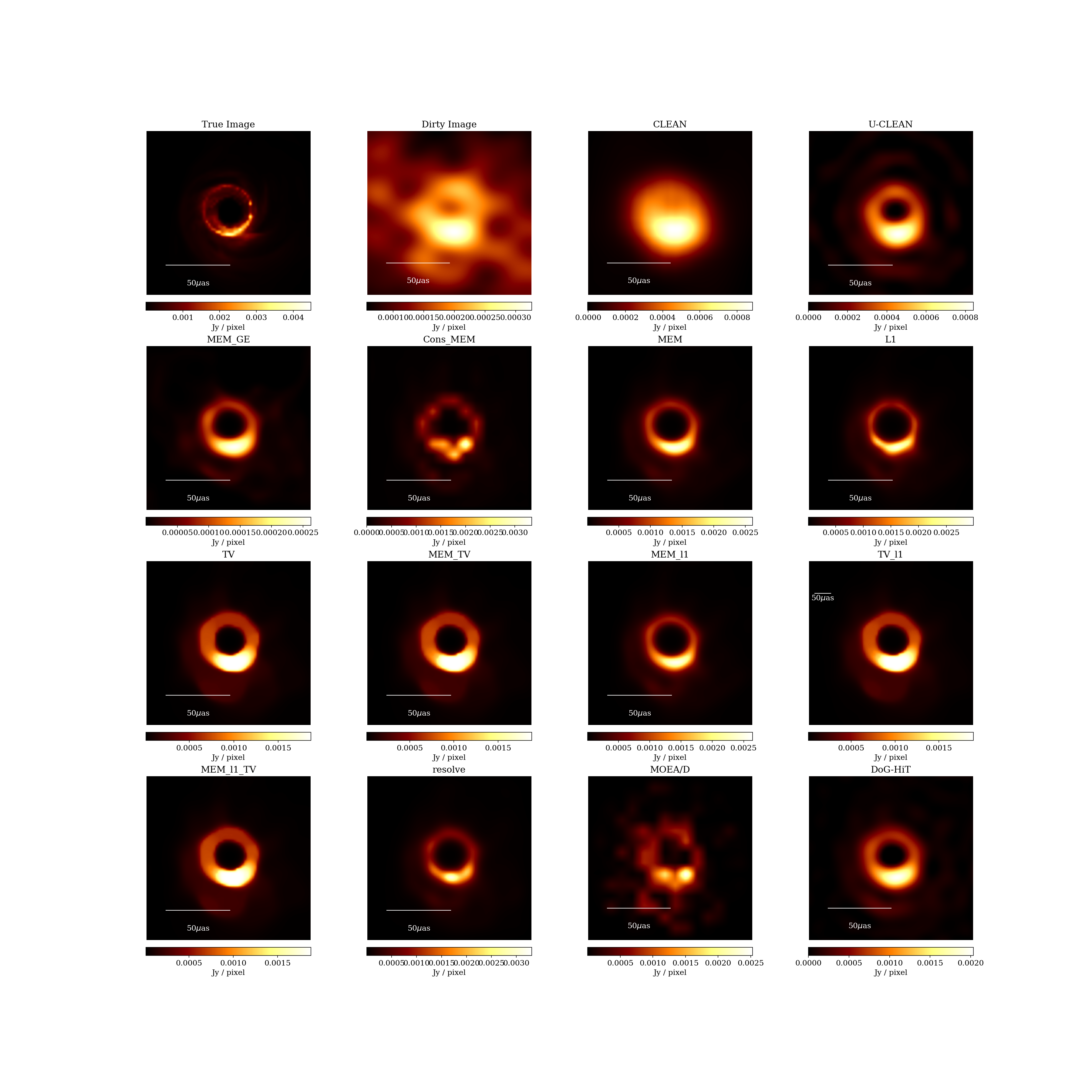}
        \caption{Reconstructions}
    \end{subfigure} \\
    \begin{subfigure}[b]{0.3\textwidth}
        \includegraphics[width=\textwidth]{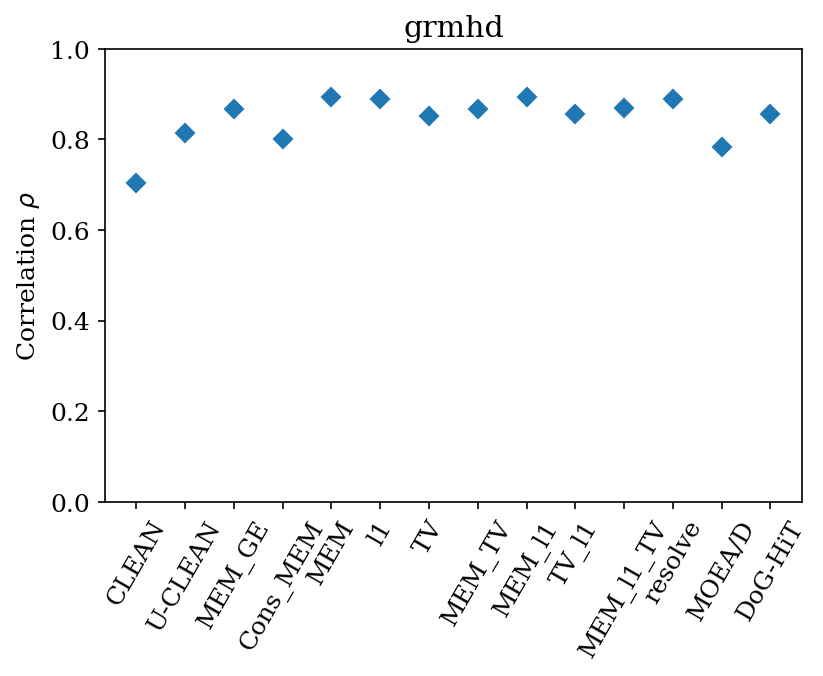}
        \caption{Correlation}
    \end{subfigure}
    \begin{subfigure}[b]{0.3\textwidth}
        \includegraphics[width=\textwidth]{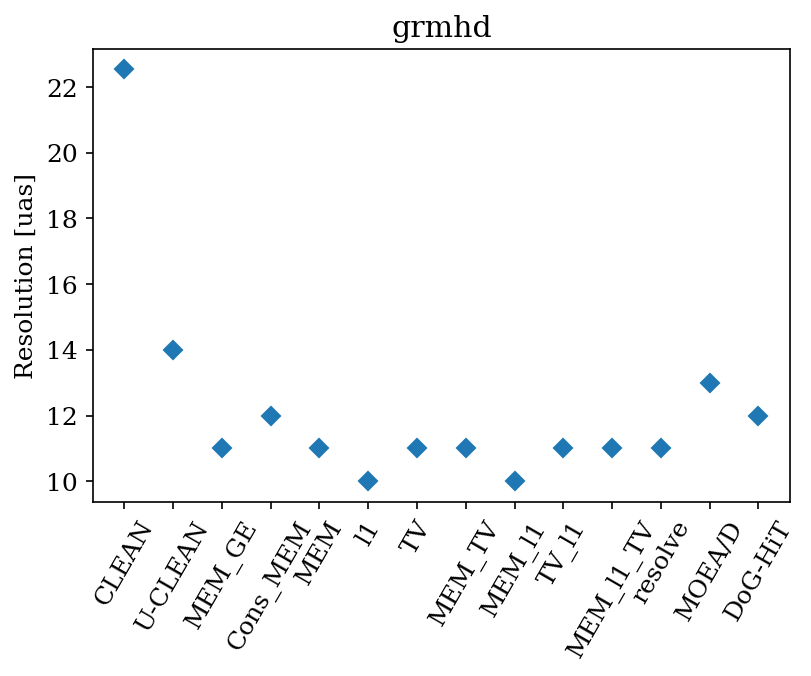}
        \caption{Resolution}
    \end{subfigure}
    \begin{subfigure}[b]{0.3\textwidth}
     \includegraphics[width=\textwidth]{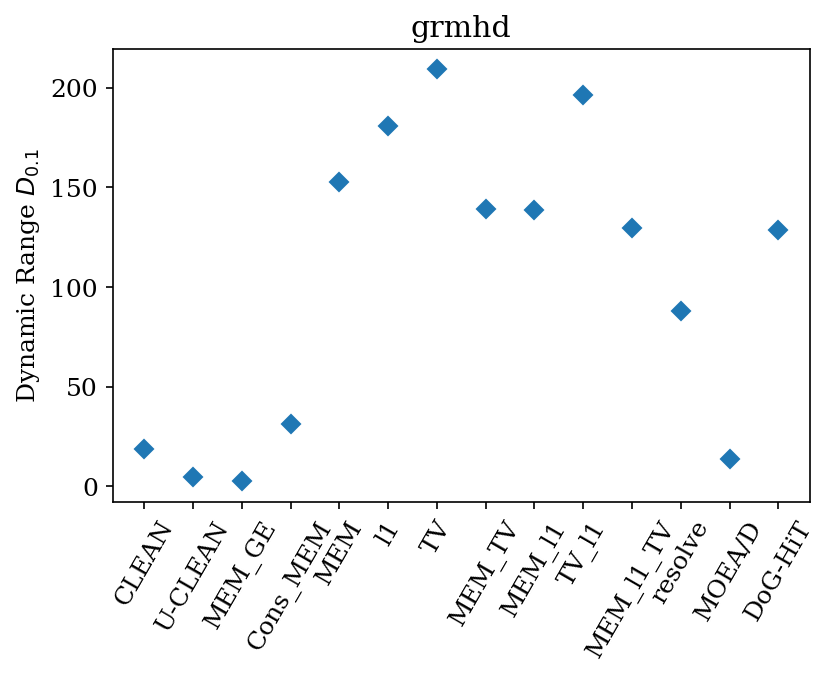}
     \caption{Dynamic Range}
    \end{subfigure}
    \caption{Reconstruction results and scoring of the various imaging algorithms on the GRMHD source model observed with the ngEHT configuration.}
    \label{fig: grmhd}
\end{figure*}

The data sets were independently analyzed in a semi-blind way with the algorithm presented in Sec. \ref{sec: imaging}. The reconstruction results are shown in Figs. \ref{fig: double_gauss_medium_stat}--\ref{fig: grmhd}. Below we provide a more detailed description of the performances of the methods, grouping them in the different categories.

\begin{figure*}
    \centering
    \begin{subfigure}[b]{0.3\textwidth}
        \includegraphics[width=\textwidth]{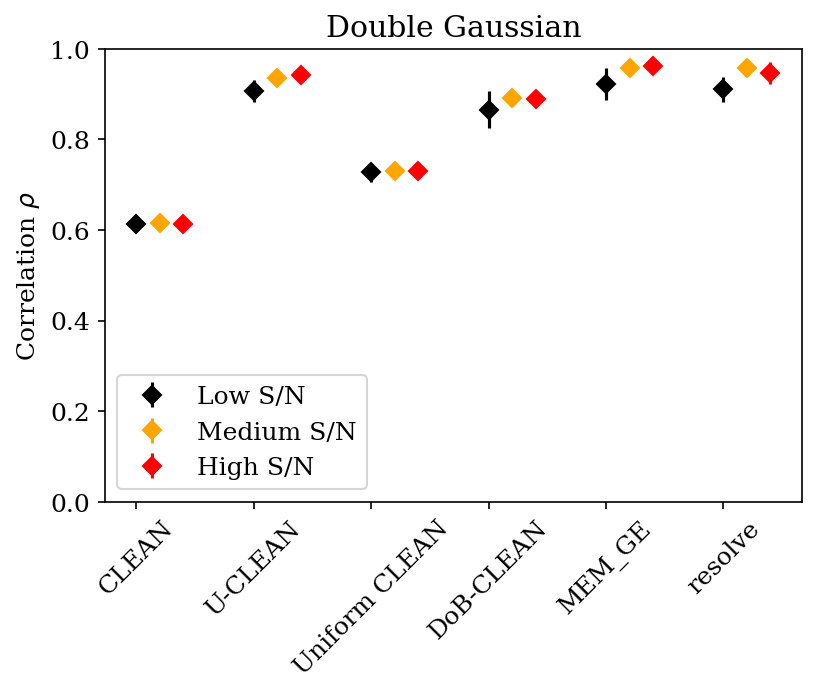}
        \caption{Correlation}
    \end{subfigure}
    \begin{subfigure}[b]{0.3\textwidth}
        \includegraphics[width=\textwidth]{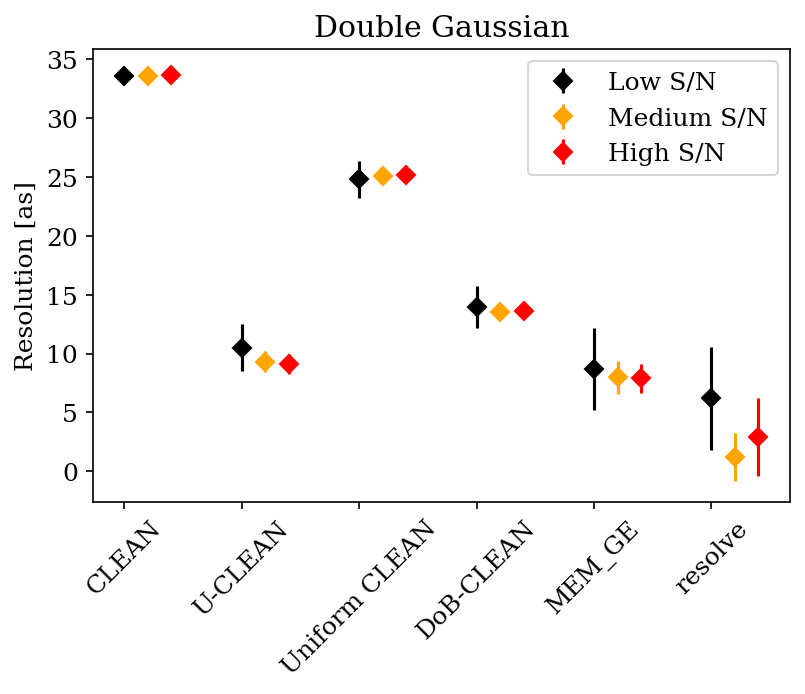}
        \caption{Resolution}
    \end{subfigure}
    \begin{subfigure}[b]{0.3\textwidth}
     \includegraphics[width=\textwidth]{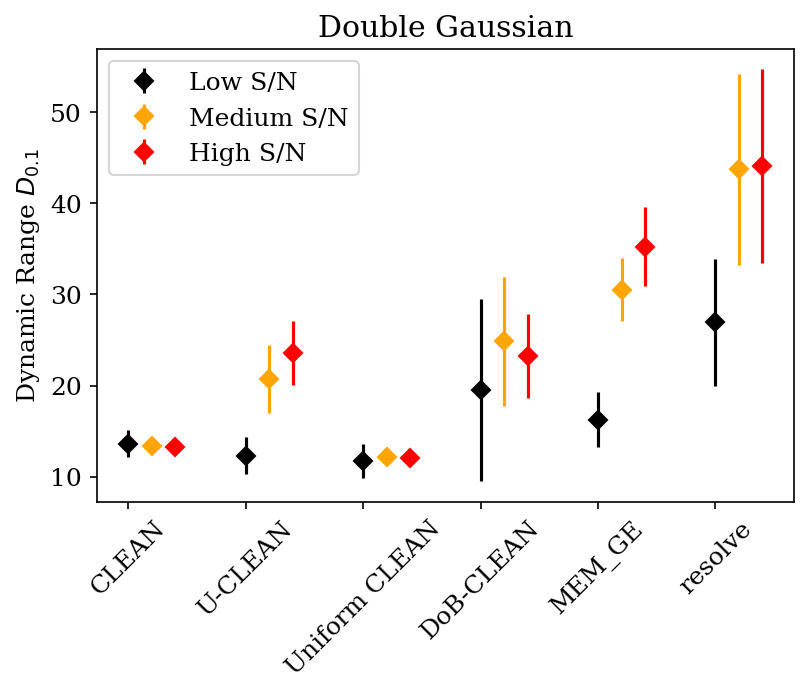}
     \caption{Dynamic Range}
    \end{subfigure}
    \caption{Comparison of the reconstruction quality among different CLEAN variants (CLEAN, U-CLEAN, Clean with uniform weights, and DoB-CLEAN), benchmarked against bayesian reconstructions and reconstructions done with MEM\_GE, in the case of the STIX double Gaussian configuration.
    }
    \label{fig: cleans}
\end{figure*}

\begin{itemize}

\item \textit{CLEAN-type algorithms.} 
CLEAN performs worse in terms of accuracy compared to all the other algorithms considered in the challenge, as proved by the systematically lowest correlation values (see panel (b) of Figs. \ref{fig: double_gauss_medium_stat}--\ref{fig: grmhd} and panel (a) of Fig. \ref{fig: cleans}). When inspecting the achieved spatial resolutions (panel (c) of Figs. \ref{fig: double_gauss_medium_stat}--\ref{fig: grmhd} and panel (b) of Fig. \ref{fig: cleans}), it gets clear that the worse performance of CLEAN is directly related to its suboptimal resolution.  While CLEAN has a well-defined resolution limit (determined by the central lobe of respective point-spread function), it has been recognized both in VLBI \citep[e.g.][]{Lobanov2005, Honma2014} and for STIX \citep[e.g.][]{massa2022hard,Perracchione2023} that this limit is too conservative in the presence of strong prior information. \citet{Lobanov2005} provide an analytic proxy for the resolution limit of an interferometric observation. While this resolution is only achievable in a specific model-fitting setting, i.e. when constraining the possible source features to Gaussian model components, it demonstrates that more sophisticated regularization methods may enable for super-resolution imaging. As we will discuss below, this is in particular achieved for the entropy-based methods (MEM\_GE, MEM, Cons\_MEM), sparsity promoting algorithms (l1) and the Bayesian reconstructions (\textit{resolve}). 
% However, in some instances (i.e. double images), Cons\_MEM and l1 reconstructions over-resolves the image, despite the noteworthy high dynamic ranges for latter one.

There are several available extensions of CLEAN that allow for super-resolution imaging. Classically, the issue of resolution is addressed by varying the CLEAN weights associated to the visibilities; we refer to \citet{Briggs1995} for a complete overview. Moreover, recently novel CLEAN variants were proposed both for STIX and VLBI that achieve super-resolution by solving the disparity between the image and the model, i.e. by making the unphysical convolution with the beam unnecessary, e.g. DoB-CLEAN \citep{Mueller2022b} and U-CLEAN \citep{Perracchione2023}. U-CLEAN is included in the overall comparison, and performs very well compared to CLEAN by outperforming CLEAN in terms of resolution, accuracy and dynamic range for all source models and experimental configurations. 

In Fig. \ref{fig: cleans}, we compare the performance of various CLEAN methods (CLEAN with natural weighting, CLEAN with uniform weighting, U-CLEAN and DoB-CLEAN) in more detail for the double Gaussian source in the case of the low, medium and high S/N.
For benchmarking, we also show the reconstruction quality of two of the best performing algorithms again, \textit{resolve} and MEM\_GE. Changing the weighting scheme is improving the scoring of the CLEAN algorithm. However, this standard and often performed, but rather simple, trick does not bring the same amount of improvements that more sophisticated interferences (such as those realized within U-CLEAN and DoB-CLEAN) offer. U-CLEAN achieves slightly higher resolutions than DoB-CLEAN does, while DoB-CLEAN achieves slightly higher dynamic ranges. While these approaches show significant improvements to standard CLEAN, they do not perform as good as the best forward modelling approaches.

CLEAN remains the de-facto standard method for VLBI and STIX reconstructions, partly due to its speed. The Fourier transform only needs to be evaluated in the major cycles, while the minor cycles only comprise of fast array substitution operations. As a consequence of the small number of pixels and the relatively simple source structures, the CLEAN reconstructions had a numerical running time of approximately 30 seconds on a standard notebook. However, CLEAN lacks a strictly defined stopping rule. Hence, the exact time of execution depends on the manually fixed number of iterations. DoB-CLEAN and U-CLEAN on the other hand incorporate more complex operations with extended basis functions. This slows the data analysis down considerably taking up to 15 minutes for DoB-CLEAN for STIX data sets. It should be mentioned here that a considerable time when using CLEAN-like algorithms is devoted to interactive choices done by the astronomer, most importantly the selection of the CLEAN windows. For the examples studied in this work this considerable effort is waived since the ground truth models are compact and rather simple, and former experience with disk masks as explored by the EHT \citep{eht2019d} existed and were utilized. The reconstruction in a completely blind, more complex data-set may take considerably longer.\\

\item \textit{Maximum Entropy Methods.} Inspecting the cross correlation metric $\rho$ for the STIX data sets, the entropy based methods perform best. There are only marginal differences between the different versions of entropy maximization, i.e. between Newton type minimization and forward-backward splitting. Cons\_MEM tops the performance overall for low SNR data, but performs less well for higher SNR data compared to MEM\_GE. This correlates with the relatively small dynamic ranges recovered for Cons\_MEM in these cases. Note that the dynamic range was computed at a common resolution (i.e. the CLEAN resolution). The performance of Cons\_MEM may be explained by the fact that regularization is less strictly employed for low SNR, since the squared programming still requires reduced $\chi^2 =1$, even for data that are highly corrupted with noise and that Cons\_MEM tends to overresolve the source structure. 
It reconstructs structures on scales smaller than the ground truth. The regularization assumption, balancing of the Lagrange multipliers, or possibly the lognormal representation of the model bias the reconstruction towards shrinked structures. This behaviour, which will be also detected for sparsity promoting RML algorithms (l1), is a warning sign to accept super-resolved structures only with relative caution, although not necessarily mirrored in the metrics presented. It is however notable that this is not an uncommon situation for imaging of sparsely sampled Fourier data in the sense that it is similar to CLEAN philosophy. It can be shown that CLEAN is effectively a sparsity promoting minimization algorithm \citep{Lannes1997} that over-resolves the image drastically, which makes a final convolution with the clean beam as a low-pass filter necessary.

MEM\_GE performs significantly better than CLEAN, but worse than RML methods for the VLBI data sets (and was more complicated to apply in practice). That is not unexpected given the much larger number of visibilities and the wide range of existing spatial scales in the image, especially for the GRMHD image. While MEM\_GE still recovers the overall structure significantly better than CLEAN, the reconstruction of the crescent observed with an EHT configuration shows some artifacts (non-closed crescent, spurious background emission that limits the dynamic range). We note that MEM\_GE is equipped with a tailored rule for the regularization parameter selection when applied to STIX data \citep[][]{Massa2020}. However, the same rule is not applicable to VLBI data and, therefore, an ad hoc choice of the regularization parameter has been adopted for reconstructing the images shown in Figs. \ref{fig: crescent} and \ref{fig: grmhd}.

We would like to highlight the remarkable performance of the plain MEM method with the crescent EHT data set in which it seems to outperform the alternative MEM\_GE approach and even Bayesian imaging. This improved performance may be attributed to the adopted form of the entropy functional. For MEM\_GE a flat prior has been used, for plain MEM a Gaussian with the size of the compact emission region. This resembles a strategy that was applied by the EHT where the size of the compact emission is constrained as an additional prior information from independent observations at smaller frequencies \citep{eht2019d}. Particularly it has been demonstrated that the size of the Gaussian prior for the computation of the entropy functional plays a significant role in the reconstruction \citep[compare the dropping percentage of top-sets for varying sizes presented in][]{eht2019d}.

Maximum entropy methods, as well as the closely related RML reconstructions, have a small numerical complexity due to the small number of visibilities and the effectivity of the limited BFGS (respectively the SQP minimization for Cons\_MEM) optimization techniques. For the STIX examples the numerical runtime took roughly a minute on a common notebook, and extends up to three minutes for the more complex ngEHT setting indicating a well scaling to a larger number of visibilities. It is further noteworthy, as mentioned above, that MEM\_GE is equipped with a tailored rule to select the regularization parameter, and Cons\_MEM is free of any tunable regularization parameters at all, marking them as remarkably fast and simple to use algorithms in practice.\\

\item \textit{Bayesian reconstruction method.} While the \textit{resolve} algorithm may be outperformed by alternative approaches in some examples, e.g. by \textit{DoG-HiT} for the EHT or by MEM\_GE in the case of STIX data, it is among the best algorithms across all instances and metrics. In particular, it shows to be an interdisciplinary alternative since it performs equally well for VLBI and STIX data analyses. Bayesian reconstructions add the additional benefit of a thorough uncertainty quantification, however at the cost of an increased complexity and computational resources; see our comparison in Appendix \ref{app: imaging}. The probabilistic approach can be beneficial for the robust image reconstruction from sparse and noisy VLBI and STIX data set. 

Bayesian methods typically need more numerical resources than comparable approaches due to the sampling. \textit{resolve} however achieves a significant speed-up by means of the adoption of variational inference (VI) methods \citep{Knollmueller2019, Frank2021}. Furthermore, we note that \textit{resolve} is the only software included in this comparison which backend has been implemented in \texttt{C++} rather than python (used for RML, MEM, \textit{DoG-HiT}, and CLEAN) or IDL (used for MEM\_GE) contributing further to a significant speed-up. For the STIX synthetic data sets, the computational time is around 2 minutes with $256 \times 256$ pixels and 6 minutes for EHT synthetic data set with $256 \times 256$ pixels. For the real STIX data sets, $512 \times 512$ pixels are used and the computational time is 6 minutes. By the standardized generative prior model in \texttt{resolve}, independent Gaussian random latent variables are mapped to the correlated log-normal distribution. The Gaussian approximated latent variables are estimated by the minimization of the Kullback-Leibler divergence in the variational inference framework. The parametrization of the latent space enables achieving affordable high-dimensional image reconstruction, although multi-modal distribution cannot be described by approximated Gaussian distribution and the uncertainty values tend to be underestimated.

For the STIX real-data application, \textit{resolve} dealt with 306197 parameters in total. \textit{resolve} depends on a number of free parameters that need to be fixed manually. For the examples studied in this manuscript, there were 11 free parameters (compared to 1 free parameter for \textit{DoG-HiT} and MEM, and 3 for RML). However, only the range (mean and standard deviation) are fixed, not the exact values to ensure flexibility on the prior.\\

\item \textit{Compressive Sensing technique.} \textit{DoG-HiT} performs overall very well, outperforming CLEAN and most non-MEM RML methods (particularly with respect to the dynamic range; see panel (d) of Figs. \ref{fig: double_gauss_medium_stat}--\ref{fig: grmhd}). However, this technique shows a slightly worse resolution induced by the nature of the extended basis functions. Particularly we would like to highlight the performance of \textit{DoG-HiT} for reconstructions with a EHT configuration, topping the performance in terms of accuracy ($\rho$) and an exceptionally high dynamic range. This is probably not surprising, since \textit{DoG-HiT} was explicitly developed for the EHT \citep{Mueller2022} and just recently saw promising application outside \citep{Mueller2023b}.

\textit{DoG-HiT} is an unsupervised and automatized variant of RML algorithms, it reduces the human bias to a minimum, making it easy and fast to apply in practice. However, the numerical resources are slightly higher. \textit{DoG-HiT} needs approximately five minutes for the reconstruction of a single STIX data set, and roughly the same time for the denser ngEHT configuration. This efficient scaling to bigger data sets is caused by the fact that while the evaluation of the Fourier transform gets numerically more expensive with a larger number of visibilities, the number of wavelets neeeded to describe the defects of the beam decreases due to the better uv-coverage as well.\\

\item \textit{Multiobjective Imaging method.} MOEA/D is the only algorithm that explores the multimodality of the problem (see Appendix \ref{app: imaging}). It finds clusters of locally optimal solutions. It computes a wide range of solutions for STIX ranging from very successful clusters comparable to entropy methods to worse reconstructions. The selection of the best cluster by the least-square principle or accumulation point criterion presented in \citet{Mueller2023} however proved challenging. When applied to STIX data, MOEA/D provides good quality reconstructions. However, it reveals to be less promising than \textit{resolve} when applied to hard X-ray visibilities. Since MOEA/D, in contrast to the other imaging algorithms described in this manuscript, does not compute a single solution, but a sample of possible image features instead, the numerical resources are comparably high. It took 90 minutes to reduce a single STIX data set, and more than four hours for the ngEHT configuration on a common notebook. There are however multiple considerations that need to be taken into account when putting these running times into context. First, the application is automatized and free of human interaction, i.e. adding no additional time for the setup of the algorithm. Second, MOEA/D works with the full set of regularizers, also including $l^2$ and total squared variation that were omitted for the RML approaches for the sake of simplicity. Lastly, a convergence analysis showed that the algorithms may have converged already after 200 rather than 1000 iterations such that the numerical running time may have been severely overestimated.\\

\item \textit{Regularized Maximum Likelihood methods.} Instead of a thorough parameter survey with all the data terms and regularization terms for RML, we study only a representative selection of terms for this manuscript inspired by the balacing principle, i.e., the scoring of all regularization terms were of similar size. In fact, the regularization terms were  chosen manually for a good performance. The impact of the regularization terms is as expected and reported in the literature. l1 promotes sparsity, thus super-resolution. TV promotes piecewise constant filaments connected by smooth functions. An entropy term promotes simplicity of the solution. For RML methods, all these terms need to be balanced properly. This leads to a high number of free hyperparameters, a severe drawback. This is typically tackled by parameter surveys, i.e. by the exploration of the scoring of the method with different weightings on synthetic data. This strategy was successfully applied in \citet{eht2019d, eht2022c} and proved convincingly the robustness of the reconstruction, but is a lengthy and time-consuming procedure. While RML methods rank among MEM methods with respect to the numerical complexity, i.e., it just took several minutes to recover a solution, the parameter survey may take significant time depending on the number of competing regularization terms that need to be surveyed. In the case of the EHT, this procedure added up to 37500 parameter combinations that needed to be surveyed \citet{eht2019d}. Without parallelization that would add up to more than 20 days of computation, but in reality always a parallel computing infrastructure is utilized. For this manuscript, we opted for a simpler approach by manually selecting well working weights by a visual inspection of keytrends on synthetic data sets. Adaptive regularization parameter updates \citep[e.g. as for MEM\_GE][]{Massa2020}, multiobjective evolution \citep[as realized for MOEA/D, see][]{Mueller2023} or the choice of more data-driven regularization terms \citep[e.g. as for \textit{DoG-HiT}, see][]{Mueller2022, Mueller2023b} were recently proposed to solve the issue of lenghty parameter surveys towards an unsupervised imaging procedure. When applied to STIX data, l1 and MEM\_l1 prove to be the best performing among all the RML methods with respect to the three metrics. In particular, they achieve dynamic range values among the highest ones within the challenge. However, a visual inspection of the reconstructions (see Figs. \ref{fig: double_gauss_medium_stat} and \ref{fig: loop_medium_stat}) shows a shrinking effect in the l1 reconstructions, in line with the sparsity promoting property of the method. This may possibly indicate an overestimated regularization parameter that led the regularization term dominate the reconstruction. The good performances of the RML methods are confirmed when applied to VLBI data, although in this case the differences between performances of the various regularization terms are less pronounced.
\\

\end{itemize}

\subsection{Real Data} \label{sec: real_data}
\begin{figure*}
    \centering
    \includegraphics[width=\textwidth]{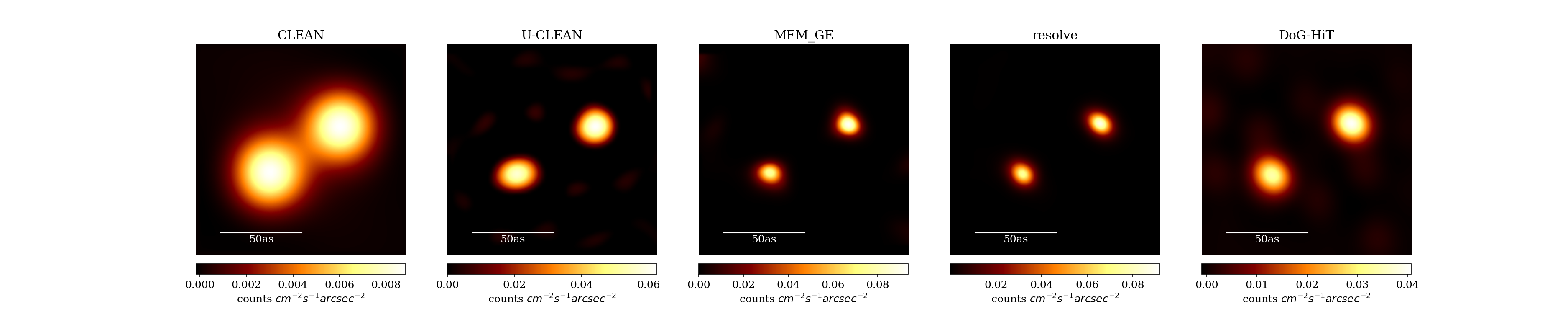}
    \caption{Imaging results of SOL2023-05-16T17 event recovered with various algorithms (CLEAN, U-CLEAN, MEM\_GE, \textit{resolve}, and \textit{DoG-HiT}, from left to right, respectively).}
    \label{fig: real_data}
\end{figure*}

As additional verification, we test selected algorithms (i.e. the best performing methods and CLEAN for benchmarking) on a real data set observed with STIX. Specifically, we consider the SOL2023-05-16T17 event integrated in the time range 17:20:30--17:21:50 and in the energy range 22-76 keV. We show the reconstruction results with five different algorithms in Fig. \ref{fig: real_data}. All algorithms successfully identify a double-gaussian structure. As discussed in the previous section in greater detail, the novel algorithms improve over CLEAN mainly by a higher resolution. All four algorithms considered here, two proposed by the STIX community, and two provided by the VLBI community, increase the resolution, and strikingly agree on the super-resolved structure.
%\Paolocomm{Would it be useful to add a table showing values of the $\chi^2$ of the different reconstructions? In this way, we could show a quantitative analysis of the goodness of the reconstructions.} \hm{I checked that all the reconstructions are fitting the data, the difference between methods arises from how the gaps are interpolated, so chi2 is not telling us anything and may be even misleading in my opinion ...}

\section{Discussion and Conclusion} \label{sec: discussion} 
In this work we explored the synergies between VLBI and STIX imaging. By cross-applying the proposed imaging algorithms and evaluating their performance, we have successfully demonstrated the strong synergies and rich opportunities of mutual interaction between the communities. This code-comparison is one of the biggest code-comparisons done in the field to date, including submissions by 17 different algorithms from a variety of algorithmic frameworks including inverse modelling, MEM, RML, 
Bayesian, multiobjective evolutionary, and compressive sensing. With the background of ongoing efforts regarding the development of novel imaging methods, we can identify some key trends that may lead future developments in the fields. Our main findings are as follows.

Modern imaging methods outperform CLEAN in terms of accuracy, dynamic range and resolution in nearly all circumstances. The amount of additional physical information that could be extracted by more sophisticated imaging algorithm is significant, and hence fosters the further development of modern imaging methods. There are long-standing and simple alternatives to the standard CLEAN procedure (e.g. varying the weighting scheme), that are in frequent use. However, significant improvements to the CLEAN strategy of imaging either requires deeper fixes to the inverse modeling pipeline as done by modern CLEAN variants \citep{Mueller2022b, Perracchione2023}, or completely changing the paradigm towards forward modeling techniques.

The VLBI side developed a wide range of RML imaging methods that deal with the multimodality of the problem \citep[MOEA/D][]{Mueller2022} and work towards unsupervised imaging with a multiscalar imaging \citep[DoG-HiT][]{Mueller2023}. While these automatized, blind imaging algorithms perform all well on STIX data sets outperforming CLEAN for a variety of data properties (i.e. noise-levels), they do not outperform the data reconstruction algorithms with significantly simpler optimization landscape, faster numerical performance and simpler use in practice, primarily MEM methods such as MEM\_GE \citep{Massa2020}. That may be an important hint for the future development of methods for STIX. Particularly, the compressive sensing algorithm \textit{DoG-HiT} does not bring the same amount of improvements as it does in VLBI, since the coverage does not allow for the same kind of separation between covered and non-covered parts.

The overall best reconstructions for STIX were achieved with entropy based algorithms (forward-backward splitting, Newton type, squared programming). Due to their additional relative simplicity, we recommend to focus the development on these methods rather than introducing the highly complex data terms recently proposed for VLBI. In this manuscript we compared three entropy-based imaging algorithms that differ in the exact form how the entropy functional is defined, the selection of the regularization parameter and the minimization procedure. They show slightly varying performance in different settings which demonstrates that the MEM approach is flexible enough to adapt to multiple situations. Bayesian imaging algorithm \textit{resolve} \citep{Arras2019, Arras2021, Arras2022} is similarly well performing both for STIX and for VLBI, and constitutes a viable alternative all over the board.

For VLBI reconstructions, the amount of available data (i.e. the sampling of the Fourier domain) and the amount of testable spatial scales (ranging several orders of magnitude) is higher. It has been demonstrated that the best results are obtained with combining many priors (l1,TV,MEM,TSV for RML, prior distribution for Bayesian) rather than with a single penalization to adapt to the fine structure \citep{eht2019d, Mueller2023}. On the contrary, we have to deal with the problems of finding the correct weightings/priors, and a misidentification may be prone to biasing the data as was observed with over-resolved structures in case of sparsity promoting regularization. The STIX algorithm MEM\_GE is strikingly successful as well, but does not yet allow for the high fidelity reconstructions that were proposed by RML or DoG-HiT specifically.

\begin{acknowledgement}
We thank the team of the ngEHT analysis challenge for providing the data set of the first ngEHT analysis challenge to us. This work was partially supported by the M2FINDERS project funded by the European Research Council (ERC) under the European Union’s Horizon 2020 Research and Innovation Programme (Grant Agreement No. 101018682) and by the MICINN Research Project PID2019-108995GB-C22. HM and JK received financial support for this research from the International Max Planck Research School (IMPRS) for Astronomy and Astrophysics at the Universities of Bonn and Cologne. AM also thanks the Generalitat Valenciana for funding, in the frame of the GenT Project CIDEGENT/2018/021. EP acknowledges the support of the Fondazione Compagnia di San Paolo within the framework of the Artificial Intelligence Call for Proposals, AIxtreme project (ID Rol: 71708).
\end{acknowledgement}

\bibliography{lib}{}
\bibliographystyle{aa}

\appendix

% \section{Imaging Methods} \label{app: imaging}
\section{Overview of the imaging methods} \label{app: imaging}
%\begin{adjustbox}{angle = 0, max height = \textheight}

\begin{table*}[h!]
    \tiny
    \centering
    \begin{tabular}{|l|l|l|} \hline
         \textbf{Method} & \textbf{(MS-)CLEAN} & \textbf{MEM} \\ \hline
         Software & Difmap, MrBeam, Casa, Aips, SSW-IDL & Casa, ehtim, MrBeam, SSW-IDL \\
         Idea & Deconvolve dirty image and dirty beam & Minimize entropy \\
         Data Term & Residual & Visibilities, Closures \\
         &  (or in basis functions for MS-CLEAN) & \\
         Minimizer/Solver & Matching Pursuit & Forward-Backward Splitting, SQP, trust-constr \\ 
         Output & Model $\neq$ image (except for DoB-CLEAN, U-CLEAN) & Regularized model \\ \hline
         Resolution& Clean beam & Super-resolution \\
         Accuracy& Small due to suboptimal representation& High \\
         Dynamic Range& High & Medium \\ \hline
         Regularization Properties& & \\
         -> Calibration& Self-calibration during imaging & Closure-only possible\\
         -> Thermal Noise& Divergence!, manual stopping & Entropy assures simplicity\\
         -> $(u,v)$-coverage& Spurious, copy covered features in gaps & Entropy\\
         &(->DoB-CLEAN, U-CLEAN: better extrapolation) & \\ \hline
         Speed& Fast & Fast \\
         Supervision& Huge human bias & Small \\
         Resources& Small, only shifts & Medium, FFT evaluated in every iteration \\
         & and subtractions performed & \\
         Adaptability& Small, not all extensions & Medium, new entropy functionals needed \\
        & could be written as a deconvolution problem & \\
         Maternity& Probed for decades, de-facto standard & Probed for decades\\ \hline \hline
         
        \textbf{Method} & \textbf{RML} & \textbf{CS} \\ \hline
        Software & ehtim, SMILI & MrBeam \\
         Idea & Generalized Tikhonov method & Sparsity promoting regularization \\
         Data Term & Visibilities, Closures & Visibilities, Closures\\
         Reg Term &  L1, L2, TV, TSV, Entropy, Flux & L1 in wavelet basis\\
         Minimizer/Solver & Newton type & Forward-Backward Splitting \\ 
         Output & Regularized model & Regularized model \\ \hline
         Resolution& Super-resolution & Super-resolution \\
         Accuracy& Highest (for correct parameter weighting) & High\\
         Dynamic Range& Medium, limited by field of view& High (multiscalar representation) \\ \hline
         Regularization Properties& & \\
         -> Calibration& Closure-only & Closure-only \\
         -> Thermal Noise& By balancing reg. terms with data terms & By balancing\\
         -> $(u,v)$-coverage& By balancing & Multiscalar Dictionary \\
         & & adapts to the $(u,v)$-coverage \\ \hline
         Speed& Fast (but parameter surveys needed) & Fast, no survey needed\\
         Supervision& Small, but parameter survey needed & Unsupervised \\
         Resources& Medium, FFT evaluated in every iteration & Medium, FFT evaluated in every iteration \\
         Adaptability&  Medium, new data terms needed & High, same multiresolution support \\
        &  &  information could be reused\\
         Maternity& Intensively tested for the EHT, & Relatively young\\
         & rare application outside & \\ \hline \hline

        \textbf{Methods} & \textbf{Bayesian} & \textbf{Multiobjective} \\ \hline
        Software & Resolve, Themis, Comrade & MrBeam\\
         Idea & Posterior exploration & Multiobjective Pareto optimality \\
         Data Term & Likelihood (Visibilities, Closures)& Closures\\
         Reg Term & Prior distribution & Multiobjective combination of\\
         & & L1,L2,TV,TSV,Entropy,Flux \\
         Minimizer/ Posterior estimation & Newton type / VI, MCMC & Genetic Algorithm \\
         Output & Posterior distribution from posterior samples & Pareto front (clusters of solutions) \\ \hline
         Resolution& Super-resolution & Super-resolved clusters\\
         & determined by averaging & as well as blurred clusters \\
         Accuracy& Highest & Limited by number of pixels\\
         & &  and genetic optimization \\
         Dynamic Range& High & Limited by number of pixels\\ 
         & & and genetic optimization \\ \hline
         Regularization Properties& &\\
         -> Calibration& Built in Bayesian model & Closure-only\\
         -> Thermal Noise& By prior distribution & By balancing multiobjective functionals \\
         -> $(u,v)$-coverage& By prior distribution & By balancing\\ \hline
         Speed& Slow & Slow, but no survey needed\\
         Supervision& Small, but larger number of parameters & Unsupervised\\
         Resources& High due to high  & High, FFT evaluated in every iteration \\
         & dimensionality of the problem & on the full population \\
         Adaptability& Medium, need to be built & Medium,
new reg. terms needed \\
        & in the prior model & \\
         Maternity& Probed in practice & In development\\ \hline
    \end{tabular}
\caption{Tabellaric overview of the properties, advantages and disadvantages of imaging frameworks that are frequently used in VLBI and for STIX.}
\label{tab: imaging}
\end{table*}
%\end{adjustbox}
\end{document}